%% file: main.tex
\documentclass[%
 reprint,
 amsmath,amssymb,
 aps,
]{revtex4-1}
\usepackage{array}
\usepackage{graphicx}
\usepackage{appendix}
\usepackage{enumitem}
\usepackage{bm}
\usepackage{amsmath}%
\usepackage{MnSymbol}%
\usepackage{wasysym}%
\usepackage{verbatim}
\usepackage{graphicx}% Include figure files
\usepackage{dcolumn}% Align table columns on decimal point
\usepackage{revsymb}
\usepackage{amsbsy}
\usepackage{amsmath}
\usepackage{tikz}
\usepackage{amsthm}
\usepackage{float}
\usepackage{natbib}
\usepackage{hyperref}
\theoremstyle{plain}
\usepackage[caption=false]{subfig}
\newtheorem{theorem}{Theorem}

\newtheorem{property}{Property}

\begin{document}

\theoremstyle{definition}

\title{ Complementarity of genuine multipartite (Bell) non-locality}

\author{Sasha Sami}
\affiliation{Center for Security, Theory and Algorithmic Research, International Institute of Information Technology, Gachibowli, Hyderabad, India.}
\author{Anubhav Chaturvedi}
\email{anubhav.chaturvedi@research.iiit.ac.in}
\affiliation{Center for Security, Theory and Algorithmic Research, International Institute of Information Technology, Gachibowli, Hyderabad, India.}
\affiliation{Institute of  Theoretical Physics and Astrophysics, National Quantum
Information Centre, Faculty of Mathematics, Physics and Informatics, University of Gda\'nsk, Wita Stwosza 57, 80-308 Gda\'nsk, Poland}
\author{Indranil Chakrabarty}
\affiliation{Center for Security, Theory and Algorithmic Research, International Institute of Information Technology, Gachibowli, Hyderabad, India.}
\begin{abstract}
We introduce a new feature of no-signaling (Bell) non-local theories, namely, when a system of multiple parties manifests genuine non-local correlation, then there cannot be arbitrarily high non-local correlation among any subset of the parties. We call this feature, \textit{complementarity of genuine multipartite non-locality}. We use Svetlichny's criterion for genuine multipartite non-locality and non-local games to derive the complementarity relations under no-signaling constraints. We find that the complementarity relations are tightened for the much stricter quantum constraints. We compare this notion with the well-known notion of \textit{monogamy of non-locality}. As a consequence, we obtain tighter non-trivial monogamy relations that take into account genuine multipartite non-locality. Furthermore, we provide numerical evidence showcasing this feature using a bipartite measure and several other well-known tripartite measures of non-locality. 
\end{abstract}

\maketitle 

\section{Introduction}
In 1964 John Stewart Bell established the fact that no physical theory of local hidden variables can reproduce all of the predictions of quantum mechanics \cite{bell1964einstein,bell1966problem,brunner2014bell}. An inequality was designed to illustrate the same. This inequality, now commonly referred to as the Bell's inequality, was based on the assumptions of local realism \cite{clauser1969proposed,freedman1972experimental}; i.e. the inequality is satisfied by all possible statistics admitted by local-real (local hidden variable) models. The phenomenon of the violation of this inequality is commonly referred to as (Bell) non-locality. The term `non-locality' is a misnomer, in this work we use the term to imply the departure of the predictions of the theory under consideration from the predictions of any local-real (local hidden variable) model.   
In quantum theory, certain states (for instance, the singlet) violate the Bell's inequality, deeming the theory to be non-local. However, the notion of non-locality is \textit{not} restricted to the quantum theory; i.e., in general the statistics predicted by any theory which violates the Bell's inequality cannot be explained by a local hidden variable model and the theory can be regarded as being non-local. 
The amount of violation of the Bell-inequality serves as a metric for non-locality; i.e., it quantifies just how far the predictions of a particular operational theory are from that of the local hidden variable models. For instance, it is known that the no-signaling condition allows more non-locality than what is admitted by the quantum theory \cite{barrett2005popescu,chakrabarty2014ctc}.
The amount of non-locality is directly associated with the advantage in many information processing protocols and tasks, when using non-local resources as compared to when using classical (local-real) resources \cite{horodecki1996teleportation,adhikari2008quantum,chakrabarty2011deletion,chaturvedi2015measurement}. \\
The notion of non-locality as presented in the Bell's theorem was restricted to bipartite systems. The extension of the notion of non-locality to multipartite systems presents itself with delicate intricacies \cite{brunner2014bell}. Let us consider a tripartite system. A well known Bell-type inequality, now commonly known as the Mermin's inequality \cite{mermin1980quantum,mermin1990extreme}, is based on the assumption that all three systems are locally correlated; i.e., the inequality is satisfied by all possible statistics of the tripartite system that can be explained by local-real models. A violation of this inequality simply implies non-locality. A bipartite non-local system locally correlated with the third system violates this inequality. The observation of this fact preceded the search for Bell-type inequalities that would distill genuine tripartite non-locality; i.e., non-locality vested in all three systems combined. Svetlichny gave a systematic way of characterizing genuine tripartite non-locality along-with an inequality, now commonly referred to as the Svetlichny's inequality \cite{svetlichny1987distinguishing}. This stronger inequality is based on the assumption of bi-locality which allows for non-locality in any bipartite subsystem locally correlated with the third system \cite{svetlichny1987distinguishing}. The violation of the Svetlichny's inequality is a sufficient condition for witnessing genuine tripartite non-locality \cite{bancal2011device}. The inequality itself is readily generalized to the multipartite scenario, wherein the Svetlichny's set of inequalities form the sufficient conditions for witnessing genuine multipartite non-locality.\\
The notion of complementarity of physical phenomenon forms an integral part of the structure of quantum mechanics. Historically, soon after the conception of the uncertainty principle \cite{heisenberg1927anschaulichen}, Neils Bohr brought up the notion of complementarity; i.e., physical systems described by the quantum theory have complementary properties which \textit{cannot} all be observed or measured simultaneously, examples include, position and momentum, energy and duration, spin on different axes, and more \cite{sazim2015complementarity}. Recently, an interesting instance was brought up in the form of complementarity between tripartite quantum correlation and bipartite Bell-inequality violation \cite{pandya2016complementarity}.\\ 
In this work, we bring forth yet another instance of complementarity, but this time around without invoking the particulars of the quantum theory. We show that in any no-signaling non-local theory, witnessing genuine $n$-party non-locality in a $n$-party system, restricts the amount of genuine $k$-party non-locality attainable by \textit{any} of its $k$-party subsystems. We use the Svetlichny's criterion for genuine multipartite non-locality and the framework of non-local games to analytically proof our main result. We further find that the relations we derive, not only hold, but are tightened under the more restrictive quantum constraints. We also investigate the relationship between the prevalent notion of monogamy of non-locality and the notion of complementarity of genuine multipartite non-locality as presented in this work. As a consequence of this investigation, we find that in any no-signaling non-local theory the monogamy relations are tightened with increase in genuine multipartite non-locality.
Furthermore, we provide numerical evidence in support of our main result, using other non-local games and inequalities in the simpler, tripartite vs. bipartite scenario. \\

\section{Preliminary concepts}
In this section, we briefly discuss the notion of non-local games, their relation to Bell-type inequalities, followed by Svetlichny's criterion of genuine multipartite non-locality. A reader familiar with these notions may consider continuing directly from the next section. 
\subsection{Non-local games}
Non-local games are convenient constructs developed primarily for showcasing the power of non-locality as a resource \cite{ambainis2013worst,brunner2014bell}.
The non-local games were designed such that their associated winning probability severed as a metric to quantify the distance between the predictions of local-real models and the quantum theory. Now they are used to compare the predictions of other models and theories as well, for instance, the bi-local model, almost quantum models and the no-signaling framework. A non-local game is a cooperative task involving two or more spatially separated parties. These games are generally played against a referee. The parties are not allowed to communicate with each other during the task. However, they are allowed to share resources (typically, any amount of classical randomness and correlations allowed by the model under consideration) before the task begins. \\ 
In this work we consider non-local games involving $n$ parties $A_1,A_2,\ldots A_n$. For each round the referee assigns the $i$th party an input bit $x_i$ and obtains an output bit $u_i$ from the $i$th party. The players win the round if their outputs satisfy a particular condition, which may depend on the assignment of the inputs in that round. The measure of success in this task is simply the probability with which the players are able to satisfy the condition.
For the rest of this paper, the winning condition is represented by a Boolean equation, say $E$, defined over the input and output bits of the parties involved in the non-local game. Suppose that, the parties share a correlation represented by a conditional probability distribution $p(u_1,u_2,\ldots u_n|x_1,x_2,\ldots x_n)$. Then the success probability of the non-local game is simply the probability with which the parties win or satisfy the equation $E$, which is given by,  
\begin{equation}
p(E)=\frac{1}{2^n}\sum_{E}p(u_1,u_2,\ldots u_n|x_1,x_2,\ldots x_n),
\end{equation}
where $\sum_{E}$ represents a sum over all inputs and outputs which satisfy $E$.\\
Each Bell-type inequality comprises of a Bell-type expression and limit on this expression. This expression is usually a linear combination of expectations of products of outputs (in $\{+1,-1\}$) of all the parties for particular input (measurement) settings. Each such Bell-type expression is inter-convertible with a non-local game, upon relabeling of outputs. To illustrate this, we shall now consider the case of two parties. 
Here we have the well known CHSH expression,
\begin{multline}
\mathcal{I}_{CHSH} = \langle x_1=0,x_2=0 \rangle + \langle x_1=1,x_2=0 \rangle \\
 + \langle x_1=0,x_2=1 \rangle - \langle x_1=1,x_2=1 \rangle.
\end{multline}
where $\langle x_1=i,x_2=j \rangle$ is simply average over outcomes $\tilde{u_1},\tilde{u_2} \in \{-1,+1\}$ of parties $A_1, A_2$ when $x_1=i,x_2=j$, i.e., 
\begin{multline}
\langle x_1=i,x_2=j \rangle= \\\sum_{k,l \in \{-1,+1\}}(k.l) p(\tilde{u}_1=k,\tilde{u}_2=l|x_1=i,x_2=j).
\end{multline}
While the Bell expressions use a `physicist' notation; i.e, the outcomes take values in $\{-1,+1\}$, non-local games utilize a `computer scientist' notation; i.e., the outcomes take values in $\{0,1\}$ \cite{pironio2011extremal}. We can rewrite $\mathcal{I}_{CHSH}$ as,
\begin{multline}\label{refmenow}
\sum_{k,l,i,j \in \{0,1\}}(-1)^{k\oplus l \oplus  ij }p(u_1=k,u_2=l|x_1=i,x_2=j).
\end{multline}
Now let us consider a non-local game described by the equation $S_2\equiv u_1 \oplus u_2 =x_1x_2$. It straightforward to see that the winning probability of this non-local game is linearly related to the value of the CHSH expression; i.e, 
\begin{equation}
p(S_2)=\sum_{S_2}p(u_1,u_2|x_1,x_2)=\frac{1}{2}\big(1+\frac{\mathcal{I}_{CHSH}}{4}\big) \;,
\end{equation}
which allows one to write the CHSH inequality in terms of the success probability of the non-local game represented by $S_2$; i.e.,
\begin{equation}\label{CHSH}
\max_{LV}\{|\mathcal{I}_{CHSH}|\} = 2 \equiv \\ \max_{LV}\{p(S_2)\}= \frac{3}{4},
\end{equation}
where $\max_{LV}$ implies maximization over all possible statistics admitted by local hidden variable (local real) models. 
\subsection{Svetlichny's criterion} 
Svetlichny's criterion for genuine mulitpartite non-locality is applicable to a system of arbitrary number of spatially separated parties. But for the sake of a brief overview, we restrict to the case of three spatially separated parties $A_1,A_2,A_3$. Each party is assigned an input bit $x_i$ and obtains and output bit $u_i$ with $i \in \{1,2,3\}$. The correlation shared by the three parties is represented by the joint probability distribution $p(u_1,u_2,u_3|x_1,x_2,x_3)$.
This correlation would adhere to a local-real model if it can be decomposed as,
%\begin{widetext}
\begin{eqnarray} \label{ML}
&&p(u_1,u_2,u_3|x_1,x_2,x_3) \nonumber \\&&= \sum_{\lambda} q(\lambda) p_{\lambda}(u_1|x_1)p_{\lambda}(u_2|x_2)p_{\lambda}(u_3|x_3),
\end{eqnarray}
 where $\lambda$ is the hidden variable and $q(\lambda)$ is the probability distribution governing $\lambda$, such that
$\sum_\lambda q(\lambda) = 1$. The correlations which \textit{cannot} be written in this way are termed as being non-local. But in the multipartite scenario, there are many distinct forms of non-locality, some more refined than others. For instance consider the case when $A_3$'s system is uncorrelated with that of $A_1$ and $A_2$,
\begin{eqnarray}
&&p(u_1,u_2,u_3|x_1,x_2,x_3) \nonumber \\ &&= p(u_1,u_2|x_1,x_2)p(u_3|x_3). 
\end{eqnarray} 
If the correlation shared by $A_1,A_2$, $p(u_1,u_2|x_1,x_2)$, violates the CHSH inequality, this deems the whole of $p(u_1,u_2,u_3|x_1,x_2,x_3)$ to be non-local. However, in this case \textit{no} non-locality, at all, is exhibited across the bi-partition, $A_1A_2,A_3$ (no correlation between $A_1,A_3$ or $A_2,A_3$). Svetlichny argued that genuine multipartite non-local correlation must have non-locality across all possible bi-partitions. Consider the following decomposition of $p(u_1,u_2,u_3|x_1,x_2,x_3)$,
%\begin{widetext}
\begin{eqnarray}\label{SL}
&&p(u_1,u_2,u_3|x_1,x_2,x_3){}\nonumber\\&& =  \sum_{\lambda_1} q_1(\lambda_1)p_{\lambda_1}(u_1,u_2|x_1,x_2)p_{\lambda_1}(u_3|x_3) {}\nonumber\\&&
           +  \sum_{\lambda_2} q_2({\lambda_2})p_{\lambda_2}(u_2,u_3|x_2,x_3)p_{\lambda_2}(u_1|x_1) {}\nonumber\\&& 
           +  \sum_{\lambda_3} q_3({\lambda_3})p_{\lambda_3}(u_1,u_3|x_1,x_3)p_{\lambda_3}(u_2|x_2) \; ,
\end{eqnarray}
%\end{widetext}
where $\lambda_1,{\lambda_2},{\lambda_3}$ are hidden variables shared by the pairs $(A_1,A_2)$,$(A_2,A_3)$ and $(A_1,A_3)$ respectively. Here, $q_1(\lambda_1),q_2({\lambda_2}),q_3({\lambda_3})$ are probability distributions over respective hidden variables, such that, $\sum_{\lambda_1} q_1(\lambda_1)+\sum_{\lambda_2} q_2({\lambda_2})+\sum_{\lambda_3} q_3({\lambda_3})=1$. \\
The right hand side of (\ref{SL}), consists of three terms, where the first allows correlation with non-locality between $A_1,A_2$, second between $A_2,A_3$ and third between $A_1,A_3$. The correlations,  $p(u_1,u_2,u_3|x_1,x_2,x_3)$ which \textit{can} be written in the form given by (\ref{SL}), are termed as being bi-local (BL). \textit{If and only if} $p(u_1,u_2,u_3|x_1,x_2,x_3)$ \textit{cannot} be written in the form given by (\ref{SL}), then we say that $A_1,A_2,A_3$ share a genuine tripartite non-local correlation or a three-way non-local correlation. It is to be noted that Svetlichny's criterion is only sufficient and that alternative criterion for genuine tripartite non-locality have also been studied \cite{bancal2011definition,gallego2012operational}. \\
\section{Svetlichny's framework}
In this section, we introduce the general form of Svetlichny's non-local games for $n$-party systems. Followed by, presentation of some essential notations, which in-turn are used to highlight certain characteristic features of Svetlichny's family of non-local games. The notations and properties are further exploited to present and derive our main result in the subsequent section.  
\subsection{Svetlichny's non-local game for $n$-party system} 
Let us consider $n$ spatially separated parties $A_1,A_2 \ldots A_n$, with input bits $x_1,x_2 \ldots x_n$ and output bits $u_1,u_2 \ldots u_n$. Then the general Svetlichny's non-local game for a $n$-party system is represented by the Boolean equation $S_n$, with the following form, 
\begin{equation}
\label{svetlichny_recurssive}
S_n \equiv  \bigoplus_{1\le i \le n} u_i \oplus c_0 = (\bigoplus_{1\le i<j \le n} x_i x_j) \oplus (\bigoplus_{1 \le i \le n}c_i x_i) \; ,
\end{equation}
where the bits $c_0,c_1,\ldots c_n$ decide the particular form of the $S_n$-game. One of the motivating factors for using Svetlichny's non-local games in this work is that, irrespective of the particular assignment of $c_k\in\{0,1\}$, the following holds for all $n\geq2$,
\begin{eqnarray}
&&\max_{BL}\{p(S_n)\}=\frac{3}{4},\nonumber \\
&&\max_{Q}\{p(S_n)\}=\frac{2+\sqrt[]{2}}{4},\nonumber \\
&&\max_{\mathcal{NS}}\{p(S_n)\}=1,
\end{eqnarray}
where $\max_{BL},\max_{Q},\max_{\mathcal{NS}}$ imply maximization over all possible correlations allowed by the bi-local models, the quantum theory and the no-signaling condition, respectively. The fact that the success probability of all $n$-party bi-local correlations in the $n$-party Svetlichny's non-local game is bounded by $\frac{3}{4}$, is referred to as the Svetlichny's inequality (the Svetlichny's inequalities also have an equivalent, rather prevalent representation with outputs in $\{+1,-1\}$)). Therefore, $p(S_n)>\frac{3}{4}$ implies genuine $n$-party non-locality.
For the two party scenario, i.e., for $n=2$ the notion of bi-locality does not make any sense, instead here the maximization of $p(S_2)$ is over the local-real models but with the same upper bound (\ref{CHSH}). For a single party, $n=1$, the local-real models, the quantum theory and the no-signaling condition allow the same maximum winning probability $\max\{p(S_1)\}=1$. 

\subsection{Observations and notations}
\begin{itemize}
\item Since there are ${n+1}$ Boolean coefficients in the expression (\ref{svetlichny_recurssive}), $c_0,c_1,\ldots c_n$, there are $2^{n+1}$ distinct $n$-party Svetlichny's games. We shall use the notation $S_n^i$ to denote a specific $S_n$ game with a specific assignment of $c_0,c_1,\ldots c_n$.
\item We denote the game which differs from $S_n^i$ only in the value assigned to $c_0$, by ${S_n^i}'$.
\item Observe that when one fixes the value of the input $x_n$ and the output $u_n$ for the $n$th party in the expression for $S_n$ given in (\ref{svetlichny_recurssive}), one obtains the expression for a particular $(n-1)$-party Svetlichny's non-local game $S_{n-1}$. We denote the particular non-local game thus obtained as $S_{n-1}^j\equiv S_n^i||(u_n=u_n',x_n=x_n')$. Where $S_n^i||(u_n=u_n',x_n=x_n'$ represents the $S_{n-1}$ game obtained from $S_n^i$ game upon fixing $(u_n=u_n',x_n=x_n')(u_n=u_n',x_n=x_n')$.
\item One can similarly obtain a $S_{k}$ game for $k<n$ upon fixing the inputs and outputs of $n-k$ parties, $S_k^j\equiv S_n^i||(u_{k+1}=u_{k+1}',u_{k+2}=u_{k+2}',\ldots u_n=u_n',x_{k+1}=x_{k+1}',x_{k+2}=x_{k+2}',\ldots,x_n=x_n')$.
\item Consider a Boolean function defined on inputs and output bits of $n-k$ parties,
\begin{equation} \label{aux1}
\mathcal{A}_{n-k}=(\bigoplus_{k<i\le n} u_i)\oplus (\bigoplus_{k<i\le n}c_ix_i), \; 
\end{equation}
and another Boolean function, defined only on the input bits of $n-k$ parties,
\begin{equation}\label{aux2}
\mathcal{B}_{n-k}=\bigoplus_{k<i \le n}x_i \; .
\end{equation}
We represent the $k$-party Svetlichny's non-local game $S_k^j$ obtained by fixing the values of $\mathcal{A}_{n-k}$ and $\mathcal{B}_{n-k}$ as, $S_k^j\equiv S_n^i||(A_n^k=a,B_n^k=b)$   
\end{itemize}
\subsection{Useful properties}
\noindent Before we proceed to the main results we study some key features (properties) of the Svetlichny's non-local games which form the basis of our main results. It is interesting to note that these features are to be attributed to the structure of Svetlichny's non-local games and are independent of any particular theory.\\
\begin{property}
\label{trivial_complementary}
Any two instances of the general $n$-party Svetlichny's non-local game (\ref{svetlichny_recurssive}), which differ only in the assignment of $c_0$ are \textit{trivially complementary} to each other; i.e., $S_n^i$ and ${S_n^i}'$ are trivially complementary to each other and the following relation holds between their success probabilities, 
\begin{equation}
p(S_n^i)+p({S_n^i}')=1.
\end{equation}
\end{property}
This property is based on the simple fact that the winning conditions of the two games in consideration are mutually exclusive.
\begin{property}
\label{not_trivial_complementary}
For different assignments of $c_i$ where $i\in\{1,2,\ldots n\}$, the general $n$-party Svetlichny's non-local game (\ref{svetlichny_recurssive}) reduces to distinct instances which are \textit{not} trivially complementary. Suppose $S_n^i$ and $S_n^j$ differ in the assignment for one or more $c_i$s, where $i \in \{1,2,\ldots n\}$, then $ S_n^j \not \equiv {S_n^i}'$. 
\end{property}
\begin{property}
\label{trivially_complementary_recurrence}
The $k$-party Svetlichny's game obtained from (\ref{svetlichny_recurssive}) by assigning specific values to $\mathcal{A}_{n-k}$ and $\mathcal{B}_{n-k}$, is \textit{trivially complementary} to the game obtained by flipping the value assigned to $\mathcal{A}_{n-k}$; i.e., $p({S_{n}^i||(\mathcal{A}_{n-k}=0,\mathcal{B}_{n-k}=b)})+p(S_{n}^i||(\mathcal{A}_{n-k}=1,\mathcal{B}_{n-k}^k=b))=1$.
\end{property}
Observe that the $S_{n-1}$ game obtained from (\ref{svetlichny_recurssive}) after fixing the input $x_n$ and output $u_n$ of $n$th party is \textit{trivially complementary} to the $S_{n-1}$ game obtained after flipping the value assigned to the output $u_n$ while keeping the input same. Now this holds for the $n-k$ party games as well, i.e., flipping the value assigned to \textit{any} of the outputs of the $n-k$ parties while keeping the inputs same, results in \textit{trivially complementary} $S_{n-k}$ games. Since function $\mathcal{A}_{n-k}$ is dependent on the outputs of the $n-k$ parties, this fact conveniently captured in the above property.
\begin{property}
\label{fixing_one_fixes_another}
The $S_k$ games obtained by flipping the values of $\mathcal{B}_{n-k}$ are \textit{not} trivially complementary; i.e. $S_k^1\equiv S_n^i||(\mathcal{A}_{n-k}=b,\mathcal{B}_{n-k}=0)$ and $S_k^2\equiv S_n^i||(\mathcal{A}_{n-k}=b^{'},\mathcal{B}_{n-k}=1)$ are \textit{not} trivially complementary to each other; i.e., $S_k^1 \not \equiv {S_k^2}'$ . 
\end{property}
Observe that on flipping the value of the input $x_n$, the two distinct $S_{n-1}$ games, thus obtained from (\ref{svetlichny_recurssive}) are \textit{not} trivially complementary. Similarly, the above holds for the $S_k$ games obtained on flipping the value assigned to \textit{any} of the inputs of the $n-k$ parties. Interestingly, we find that there are only two distinct, \textit{not} trivially complimentary $S_k$ games possible, which basically \textit{only} depends on the particular assignment of $\mathcal{B}_{n-k}$, a function which only depends on the input bits of $n-k$ parties. However, the two games, $S_k^1$ and $S_k^2$, thus obtained, are related to each other see \textbf{Appendix B.}. 
\section{Complementarity relations}
In this section we showcase our main results. We consider $n$ spatially separated parties $A_1,A_2, \ldots A_n$, each with input bits $x_1,x_2, \ldots x_n$ and output bit $u_1,u_2, \ldots u_n$. They share a no-signaling correlation $p(u_1,u_2, \ldots u_n|x_1,x_2 \ldots x_n )$. Suppose that this correlation wins a particular Svetlichny's non-local game $S_n^i$ with a success probability $p(S_n^i)$. Then, we show that the success probability attainable by this correlation in any other $n$-party Svetlichny game $p(S_n^j)$ is upper-bounded. Furthermore, the upper-bound, $\max_{\mathcal{NS}}\{p(S_n^j)\}$ decreases with increase in $p(S_n^i)$.  
\begin{theorem}
\label{thm_1} \textit{$S_n^i$ vs. $S_n^j$}: For any $n$-party no-signaling correlation, there holds the following complementary relation between the attainable success probabilities of two distinct $n$ party Svetlichny's non-local games $S_n^i$ and $S_n^j$,
\begin{eqnarray}
\label{snsn}
\max_{\mathcal{NS}}\{p(S^j_n)\}=
      \frac{3}{2}-p({S_n^i}),
\end{eqnarray} 
where $S_n^j \not \equiv {S_n^i}'$ and $p(S_n^i)\ge \frac{1}{2}$.
\end{theorem}

\begin{figure}[hbtp] 
\includegraphics[scale=0.6]{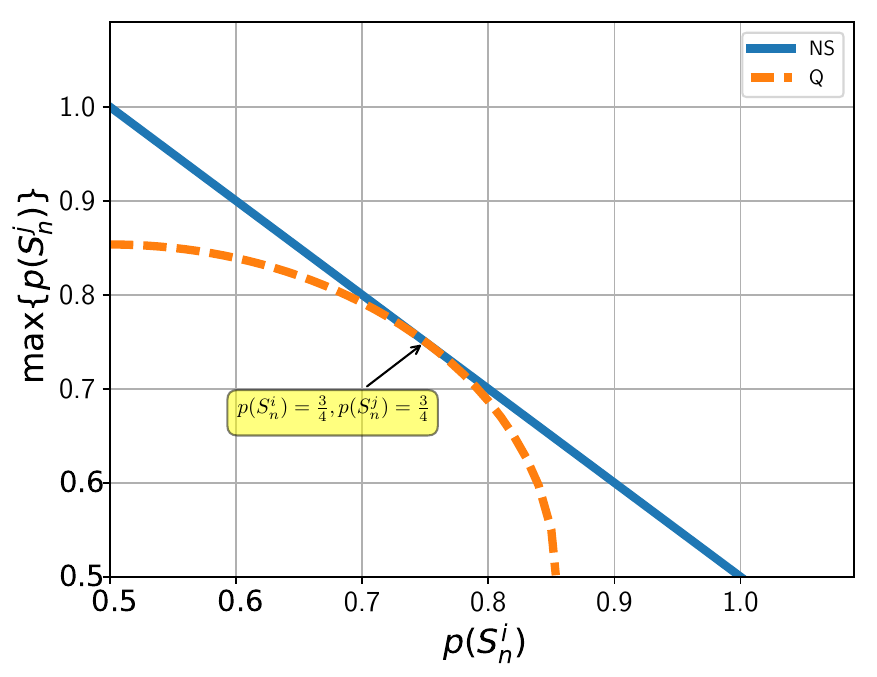}
\caption{ [color online] Plot of maximum achievable winning probability in a particular $S_n^j$ game, $\max_{\mathcal{NS}}\{p(S_n^j)\}$, against the observed winning probability in another $S_n^i$ game, $p(S_n^i)$  (when $S_n^i,S_n^j$ are \textit{not} trivially complementary). For maximization over no-signaling probability distributions, the solid blue line, and for quantum probabilities, the dashed orange curve.}
\label{snvssn} 
\end{figure}

The relations presented in this Theorem are tight. The above relation is saturated by the convex combination of two distinct Svetlichny's boxes (a no-signaling correlation that wins the Svetlichny's non-local game with unit probability), specifically $S_n^j$-box and $S_n^i$-box [see FIG. \ref{snvssn} blue curve]. These boxes also form the extremal points of the $n$-party no-signaling polytope. When $p(S^j_n)=1$, it implies that the correlation is a $S_n^j$-box with $p(S_n^i)=\frac{1}{2}$. When $p(S^i_n)=1$, it implies the correlation is a $S_n^i$-box with $p(S_n^j)=\frac{1}{2}$. It is interesting to note that it is enough to consider $p(S_n^i)\ge \frac{1}{2}$ as $p(S_n^i)=1-p({S_n^i}')$; i.e., when $p(S_n^i) < \frac{1}{2}$, then $p({S_n^i}')\ge \frac{1}{2}$ and $\max_{\mathcal{NS}}\{p(S_n^j)\}=\frac{3}{2}-p({S_n^i}')$.
For detailed proof of this Theorem see \textbf{Appendix C.}.
To approximate $\max_{Q} \{p({S_n^j})\}$, we find $\max_{Q_{n-1}} \{p({S_n^j})\}$ given an observed value of $p(S_n^i)$, where $Q_{n-1}$ is an convex set of correlations which approximates the quantum set of correlations. The test of whether a correlation belongs to this set or not can be formulated as a semi-definite program (SDP). Such SDPs form an ordered family, commonly referred to as the Navascues-Pironio-Acin (NPA) hierarchy \cite{navascues2008convergent,wehner2006tsirelson,vandenberghe1996semidefinite,liang2007bounds,doherty2008quantum}. The limit of the sequence of the ever tightening convex sets $Q_1 \supset Q_2 \supset Q_3 \ldots \supset Q_{m}$ is $\lim_{ n \Rightarrow \infty}Q_m=Q$, therefore $\max_{Q} \{p({S_n^j})\} \le \max_{Q_{n-1}} \{p({S_n^j})\}$. To plot
 $\max_{Q_{n-1}} \{p({S_n^j})\}$ against $p({S_n^i})$, we solve a SDP for $n\in \{2,3,4,5\}$ and obtain \textit{identical} curves [see FIG. \ref{snvssn} orange curve]. It is interesting to note that, $\max_{\mathcal{NS}}\{p(S_n^j)+p(S_n^i)\}=\max_{Q}\{p(S_n^j)+p(S_n^i)\}=\frac{3}{2}$, i.e., while the quantum curve always remains within the no-signaling curve they meet at a unique point where $p(S_n^j)=p(S_n^i)=\frac{3}{4}$ [this point is highlighted in FIG. \ref{snvssn}].\\
The Theorem presented above is a stepping stone to our main result presented below. Again, $n$ spatially separated parties share a no-signaling correlation $p(u_1,u_2,\ldots u_n|x_1,x_2,\ldots x_n)$. Now if using this correlation the parties can win a $n$ party Svetlichny's non-local game with a success probability $p(S_n^i)>\frac{3}{4}$, then the attainable success probability in any $k$-party Svetlichny's non-local game $p(S_k^j)$ is upper bounded, where $k<n$. Moreover, the upper bound decreases with the increase in $p(S_n^i)$.
\begin{theorem} $S_k$ vs. $S_n$:
\label{thm_2} For any $n$-party no-signaling correlation, there holds the following complementarity relation between the attainable winning probability in any $k$-party Svetlichny's non-local game $S_k^j$, where $k<n$, and the observed winning probability of $n$ party Svetlichny's non-local game $S_n^i$, given by,
\begin{eqnarray}
\label{nvn}
  \max_{\mathcal{NS}}\{p({S_{k}^j})\} =
    \begin{cases}
       \bigg( \frac{5-4p(S_{n}^i)}{2} \bigg),  \text{if}\ p(S_{n}^i) \geq \frac{3}{4} , \\
      1, \;\;\;\;\;\;\;\;\;\; \text{if}\ \frac{1}{2}\leq  p(S_{n}^i) \leq \frac{3}{4},\\
    \end{cases}{} 
\end{eqnarray} 
where $S_k^j \equiv S_n^i||(\mathcal{A}_{n-k}=a,\mathcal{B}_{n-k}=b)$ for some $a,b \in \{0,1\}$.
\end{theorem}
\begin{figure}[hbtp]
\includegraphics[scale=0.6]{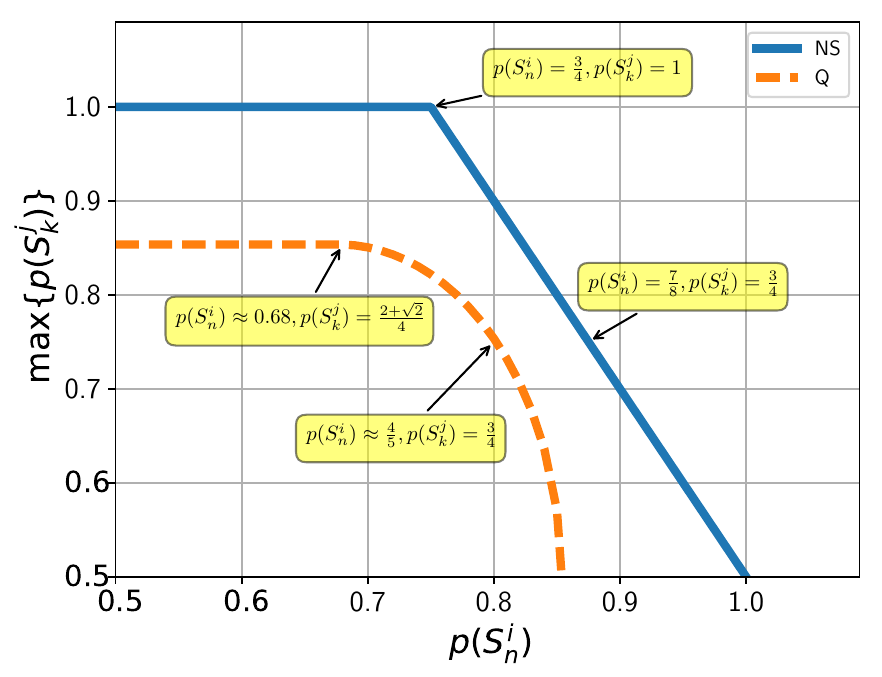}
\caption{[color online] Plot of maximum achievable winning probability in a particular $S_k^j$ game, $\max \{p(S_k^j)\}$,  against the observed winning probability in a $S_n^i$ game, $p(S_n^i)$ (when $S_k^j \equiv S_n^i|(\mathcal{A}_{n-k}=a,\mathcal{B}_{n-k}=b)$ for some $a,b \in \{0,1\}$). For maximization over no-signaling probability distributions, the blue curve, and for quantum probabilities, the orange curve.
}
\label{skvssn}
\end{figure}
The Theorem states that if one witnesses genuine $n$-party non-locality, i.e., $p(S_n) > \frac{3}{4}$, then the no-signaling constraints imply an upper bound on the amount of attainable genuine $k$-party non-locality $p(S_k)$ for any $k<n$.  
The relation presented in \textbf{Theorem 2.} is tight when the $k$-party game $S_k^j$ whose winning probability we are maximizing is one of the four $S_k$-games obtained by fixing $\mathcal{A}_{n-k},\mathcal{B}_{n-k}$ from the $n$ party $S_n^i$-game. When $p(S_k^j)=1$ we have a $S_k^j$-box. If we have a completely random distribution for the remaining $n-k$ parties, i.e., $p(u_{k+1},u_{k+2}, \ldots u_n|x_{k+1},x_{k+2}, \ldots x_n)=\frac{1}{2^{n-k}}$, we have $p(S_n^i)=\frac{1}{2}$. However, only when the $n-k$ parties have a particular deterministic distribution along with a $S_k^j$-box, $p(S_n^i)=\frac{3}{4}$. For $p(S_n^i)\ge \frac{3}{4}$, the curve in the relation is simply the convex combination between $S_k^j$-box along-with the deterministic distribution for $n-k$ parties and the $S_n^i$-box [see FIG. \ref{skvssn} blue curve]. We use the set, $Q_{n-1}$ from NPA hierarchy, to approximate the quantum case for $S_k$ vs. $S_n$. We plot  $\max_{Q_{n-1}}p(S_k^j)$ against $p(S_n^i)$ for $n\in \{2,3,4,5\}$ and $k\in \{1,2, \ldots n-1\}$ and find identical curves [see FIG. \ref{skvssn} orange curve]. In the quantum scenario, we find that the complementarity relations are tightened. It is interesting to note that the quantum curve always lies within the no-signaling curve and never touches it. As noted above, in the no-signaling case $\max_{\mathcal{NS}}\{p(S_k^j)\}$ decreases for $p(S_n^i) > \frac{3}{4}$, on the other hand in the quantum case $\max_Q\{p(S_k^j)\}$ decreases for $p(S_n^i) \gtrsim 0.68$ [these points are highlighted in FIG. \ref{skvssn}].\\    
The proof of the above Theorem relies on the fact that, upon fixing the values of  $n-k$ party Boolean variables, $\mathcal{A}_{n-k}$ and $\mathcal{B}_{n-k}$, a $S_n^i$ game yields four distinct $S_k$ games. Using \textbf{Property 3.} and \textbf{Property 4.} one may write the winning probability of a $S_n^i$ game as,
\begin{eqnarray}
\label{general_win_probability2}
&&p({S_n^i}) ={}\nonumber\\&& \frac{1}{2} \bigg( p(\mathcal{A}_{n-k}=0|\mathcal{B}_{n-k}=0)p({S_k^1}|\mathcal{A}_{n-k}=0,\mathcal{B}_{n-k}=0){}\nonumber\\&&+p(\mathcal{A}_{n-k}=1|\mathcal{B}_{n-k}=0)p({S_k^1}'|\mathcal{A}_{n-k}=1|\mathcal{B}_{n-k}=0) \nonumber\\&&
+p(\mathcal{A}_{n-k}=0|\mathcal{B}_{n-k}=1)p({S_k^2}|\mathcal{A}_{n-k}=0,\mathcal{B}_{n-k}=1){}\nonumber\\&&+p(\mathcal{A}_{n-k}=1|\mathcal{B}_{n-k}=1)p({S_k^2}'|\mathcal{A}_{n-k}=1,\mathcal{B}_{n-k}=1) \bigg), \;{}\nonumber\\&& 
\end{eqnarray}
where the conditional probability distribution $p(\mathcal{A}_{n-k}|\mathcal{B}_{n-k})$ can be computed from the marginal probability distribution $p(u_{k+1},u_{k+2} \ldots u_n|x_{k+1},x_{k+2} \ldots x_n)$ of $n-k$ parities. Now, the above relation is obtained using the relations in \textbf{Theorem 1.}; i.e., by comparing $p(S_k^1|\mathcal{A}_{n-k}=0,\mathcal{B}_{n-k}=0)$, $p({S_k^1}|\mathcal{A}_{n-k}=1,\mathcal{B}_{n-k}=0)$, $p(S_k^2|\mathcal{A}_{n-k}=0,\mathcal{B}_{n-k}=1)$ and $p({S_k^2}'|\mathcal{A}_{n-k}=1,\mathcal{B}_{n-k}=1)$ with the winning probability of a particular $k$-party Svetlichny's non-local game, say $p(S_k^1)$, irrespective of the assignment of $\mathcal{A}_{n-k},\mathcal{B}_{n-k}$ [see FIG. \ref{prove_me}]. The actual proof of this theorem is rather arduous and can be found in \textbf{Appendix E.}.\\ 
\begin{figure}[hbtp] 
\includegraphics[scale=0.45]{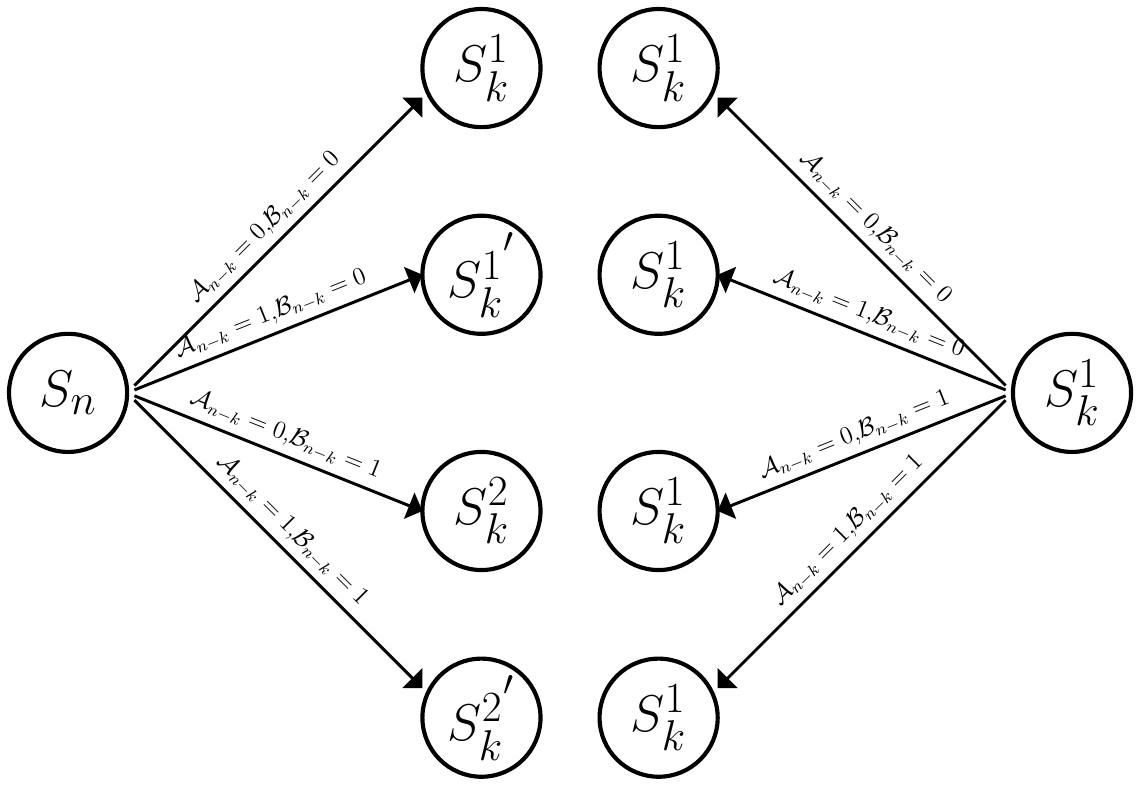}

\caption{The proof sketch for \textbf{Theorem 2.}. It utilizes the \textbf{Property 3.}, \textbf{Property 4.} and the relation presented in \textbf{Theorem 1.}.
\label{prove_me}
}
\end{figure}
\section{Relation with Monogamy of non-locality}
\begin{figure*}
\subfloat[Monogamy of non-locality.]{%
  \includegraphics[scale=0.425]{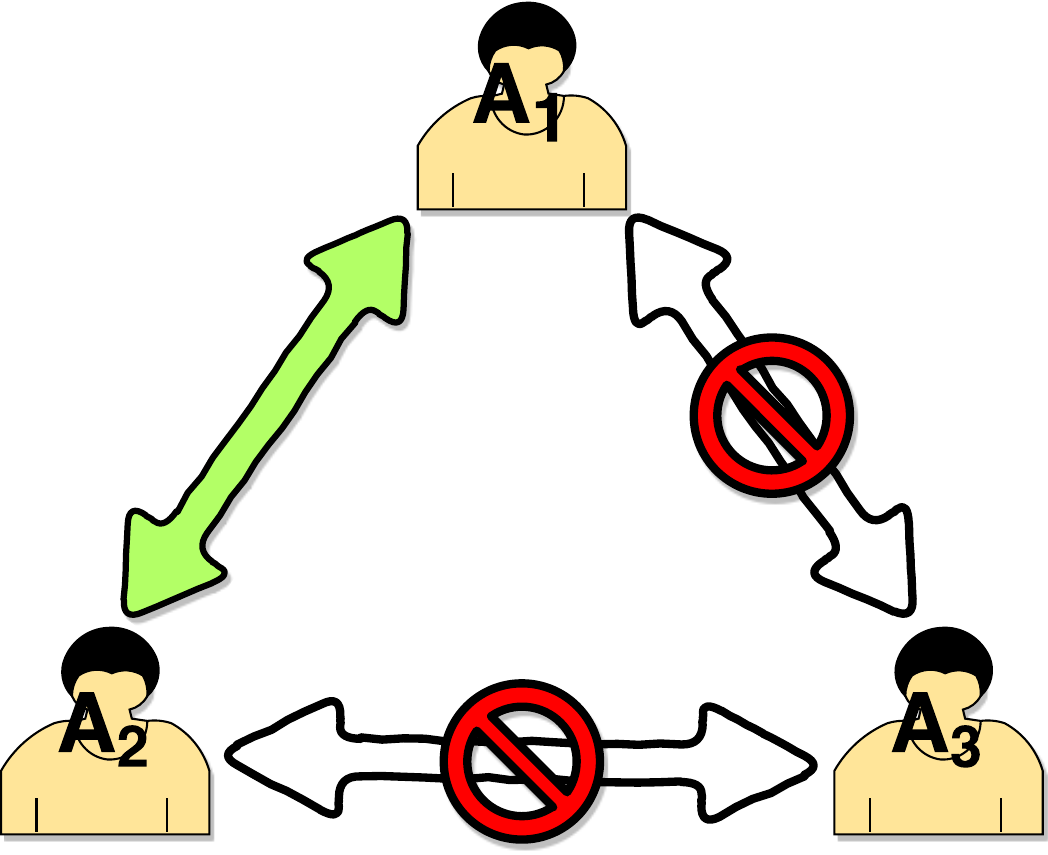}%
}\hspace{0.08\linewidth}
\subfloat[Complementarity of genuine tripartite and bipartite non-locality.]{%
  \includegraphics[scale=0.425]{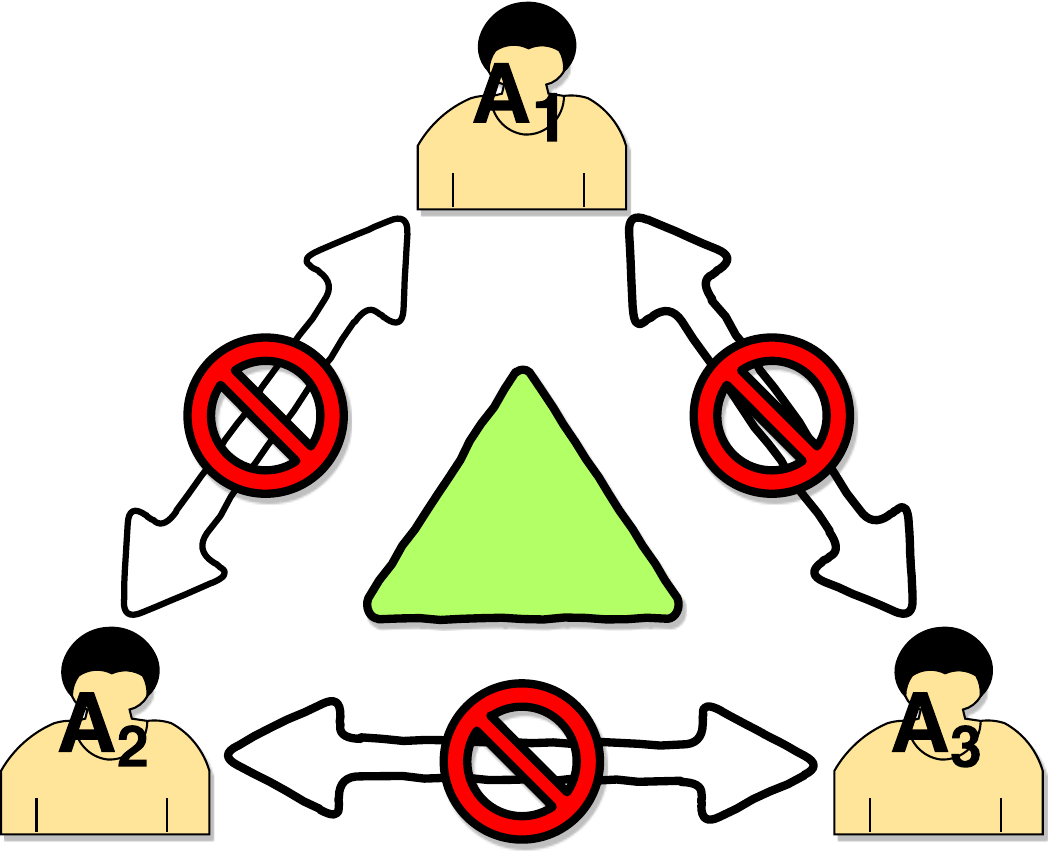}%

}
\caption{\label{idea}[color online] Extremal form of monogamy of non-locality vs. extremal form of complimentarity of genuine non-locality in a tripartite scenario.}

\end{figure*}

\begin{figure*}
\subfloat[\label{love} No-signaling]{%
  \includegraphics[scale=0.6]{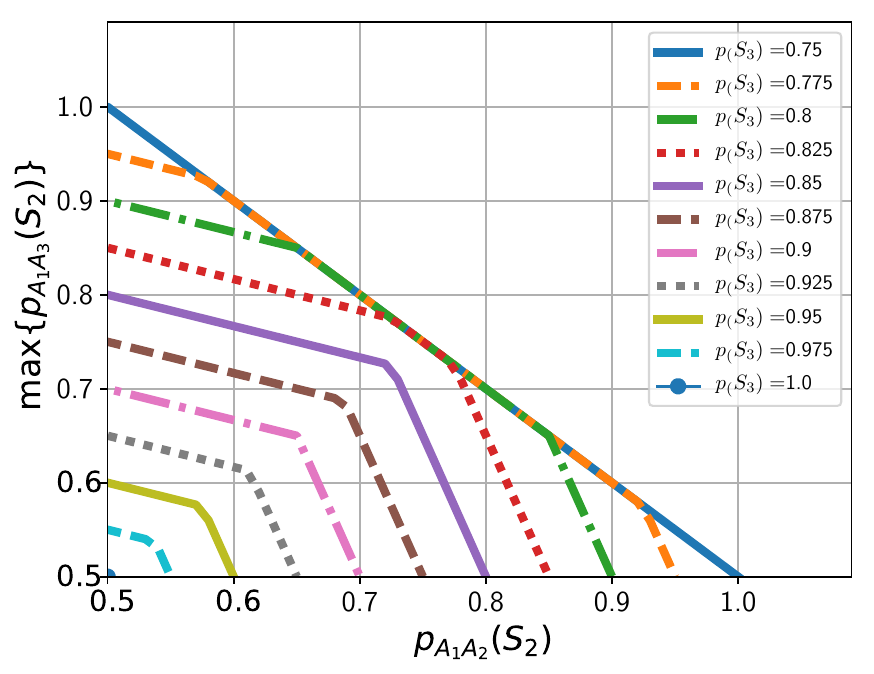}%
} \hfill
\subfloat[\label{love1} Quantum]{%
  \includegraphics[scale=0.6]{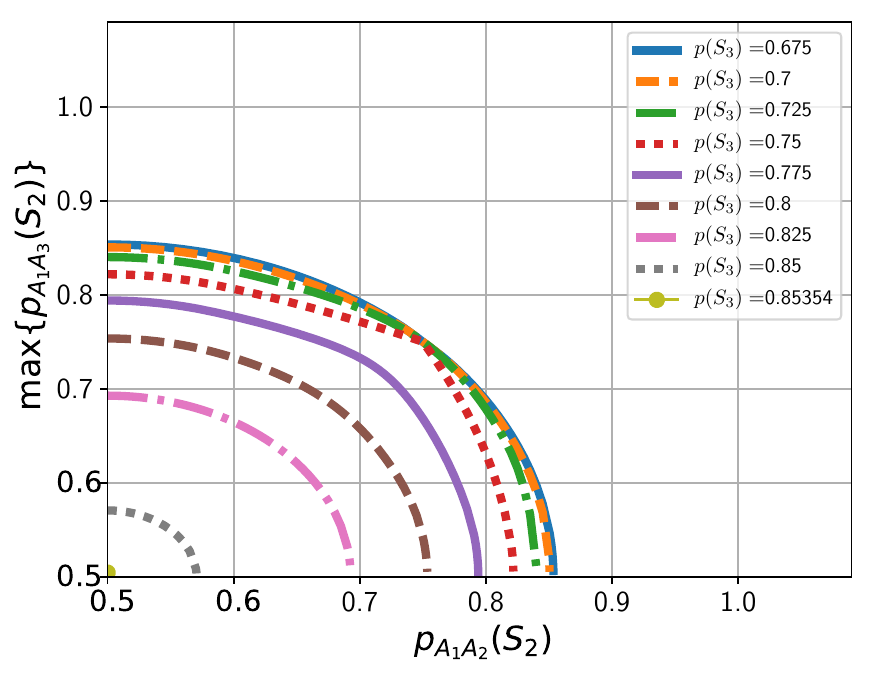}%
}
\caption{[color online] The curves of maximum achievable success probability in a $S_2$ game, played between $A_1,A_2$, $\max\{p_{A_1,A_2}(S_2)\}$, against the winning probability in a $S_2$ game, played between $A_1,A_3$, $p_{A_1,A_3}(S_2)$, given different observed value of the winning probability in a $S_3$ game, $p_{A_1,A_2,A_3}(S_3)$. The first curve from the top corresponds to the top most legend entry, the second curve to the second legend entry and so on.}
\label{ahCommon}
\end{figure*}

In this section, we highlight the key differences and the subtle relationship between monogamy of non-local correlations and complementarity of genuine multipartite non-locality. It has been observed by several authors \cite{toner2009monogamy,san2009entanglement,pawlowski2009monogamy,ramanathan2012generalized,bai2014general,osborne2006general,koashi2004monogamy,hiroshima2007monogamy}, that a system being (non-locally) correlated with another one, cannot have arbitrarily high non-local correlation with a third system: this has been dubbed as the `monogamous' nature of non-locality.
For instance, non-local correlations between spatially separated systems that lead to the violation of the well-known CHSH inequality (\ref{CHSH}), exhibit this phenomenon. It was later shown that all no-signaling correlations violating any Bell-type inequality exhibit monogamy \cite{pawlowski2009monogamy}. This feature of non-local correlations, imposed by the no-signaling condition reflects the monogamy of entanglement under the much stricter quantum constraints. Furthermore, this feature is significant in various cryptographic scenarios, where two parties $A_1$ and $A_2$ can verify a sufficient amount of non-locality to guarantee that their systems are not correlated with any eavesdropper's system \cite{toner2009monogamy}. \\
Consider the case of three spatially separated parties $A_1,A_2,A_3$, sharing a no-signaling correlation $p(u_1,u_2,u_3|x_1,x_2,x_3)$. 
It was first shown by Toner \cite{toner2009monogamy}, that in a Bell experiment with three spatially separated parties $A_1,A_2,A_3$, when $A_1,A_2$ observe the violation of the CHSH inequality, then the no-signaling condition imposes that there can be \textit{no} violation of the CHSH inequality between $A_1,A_3$. This is conveniently captured by the following `monogamy' relation:
\begin{eqnarray}
\label{mono}
\nonumber
\max_{\mathcal{NS}}\{p_{A_1,A_2}(S_2)+p_{A_1,A_3}(S_2)\} && = 2\max_{LV}\{p(S_2)\} \\  && =\frac{3}{2},
\end{eqnarray}
where
$p_{A_1,A_2}(S_2)=\sum_{u_3}\sum_{S_2}p(u_1,u_2,u_3|x_1,x_2,x_3)$ and $p_{A_1,A_3}(S_2)$ is similarly defined. This relation is saturated for no-signaling constraints for all possible values of $p_{A_1,A_2}(S_2),p_{A_1,A_3}(S_2)$ [see FIG. \ref{love} curve for $p(S_3)=0.75$ ]. Interestingly, the same relation holds for quantum case as well. However, for the quantum case it is saturated only for $p_{A_1,A_2}(S_2)=\frac{3}{4},p_{A_1,A_3}(S_2)=\frac{3}{4}$ [see FIG. \ref{love1} curve for $p(S_3)=0.675$] . \\
The complementarity relations for this tripartite system are,
\begin{eqnarray}
\label{nvn31}
  \max_{\mathcal{NS}}\{p_{A_1A_2}({S_{2}})\} =
    \begin{cases}
       \bigg( \frac{5-4p(S_{3})}{2} \bigg),  \text{if}\ p(S_{3}) \geq \frac{3}{4}, \\
      1, \;\;\;\;\;\;\;\;\;\; \text{if}\ \frac{1}{2}\leq  p(S_{3}^i) \leq \frac{3}{4},  \\
    \end{cases}{}
\end{eqnarray} and exactly same for $\max_{\mathcal{NS}}\{p_{A_1A_3}({S_{2}})\}$.
In general the monogamy relations \textit{do not} take into account the amount of genuine multipartite non-locality the system may have and are tight \textit{only} under the assumption that there isn't any as is the case with the relation in (\ref{mono}). For a multipartite system, the monogamy relations reveal that if a high degree of (genuine $n$-party) non-local correlation is observed in a system of $n$-parties then none of these $n$-parties can be highly correlated with any other external party. While on the other hand the complementarity relations presented in this work, on the other hand, deal with the distribution of genuine non-locality within a $n$-party system; i.e., they \textit{do not} take any external parties into consideration (see FIG. \ref{idea} for the visualization of the difference in the notions).

\begin{figure} 
\includegraphics[scale=0.6]{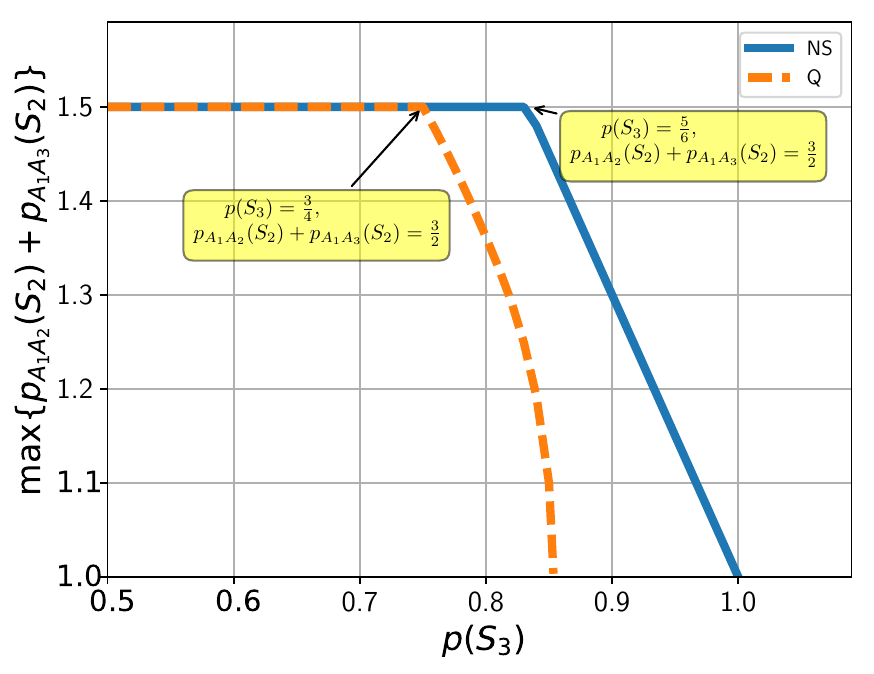}
\caption{[color online] Plot of maximum achievable value of the sum of winning probabilities of a $S_2$ game played between $A_1,A_2$ and $A_1,A_3$, $\max \{p_{A_1A_2}(S_2)+p_{A_1A_3}(S_2)\}$,  against the observed winning probability in a $S_3$ game, $p(S_3)$. For maximization over no-signaling probability distributions, the solid blue line, and for quantum probabilities, the dashed orange curve.
\label{monovssn}
}
\end{figure}

While these two features of no-signaling non-theories are distinct, they are subtly related. In particular, the monogamy relation, (\ref{mono}), tightens with increasing presence of genuine multipartite non-locality. This is captured in the following modified monogamy relation,
\begin{multline}
\label{monogamy}
  \max_{\mathcal{NS}}\{p_{A_1A_2}({S_{2}})+p_{A_1A_3}({S_{2}})\}\\  
 = \begin{cases}
       4-3p(S_{3}) ,~~~  \text{if}\ p(S_{3}) \geq \frac{5}{6}, \\
       \frac{3}{2}, ~\;\;\;\;\;\;\;\;\;~~~~~~ \text{if}\ \frac{1}{2}\leq  p(S_{3}^i) \leq \frac{5}{6}. 
    \end{cases}{}
\end{multline} 
Notice that the complementarity relations of the type (\ref{nvn31}) \textit{do not} imply the monogamy relation (\ref{monogamy}) on their own; i.e., one \textit{does not} arrive at the above result just by adding such complementarity relations for different pairs of parties, i.e., $\max_{\mathcal{NS}}\{p_{A_1A_2}(S_2)\}+\max_{\mathcal{NS}}\{p_{A_1A_3}(S_2)\} \not \equiv \max_{\mathcal{NS}}\{p_{A_1A_2}(S_2)+p_{A_1A_3}(S_2)\}$. Interestingly, the maximum value of the sum $\max_{\mathcal{NS}}\{p_{A_1A_2}(S_2)+p_{A_1A_3}(S_2)\}$ remains constant at $\frac{3}{2}$ for $\frac{1}{2} \leq p(S_3)\leq \frac{5}{6}$ and decreases for $p(S_3)>\frac{5}{6}$ [see FIG. \ref{monovssn} blue curve]. We use SDP once again to approximate the quantum case. Here the maximum value of the sum $\max_{Q}\{p_{A_1A_2}(S_2)+p_{A_1A_3}(S_2)\}$ remains constant at $\frac{3}{2}$ for $\frac{1}{2} \leq p(S_3)\leq \frac{3}{4}$ and decreases for $p(S_3)>\frac{3}{4}$ [see FIG. \ref{monovssn} orange curve].	
 The above result can easily be proved using exactly the same proof structure used for \textbf{Theorem \ref{thm_2}.}. 
\\
However, the above relation \textit{does not} capture the delicate details involved in this scenario.
Observe that in the above relation the maximal value of the sum $\max_{\mathcal{NS}}\{p_{A_1A_2}(S_2)+p_{A_1A_3}(S_2)\}$ decreases only after $p(S_3) > \frac{5}{6}$. On the other hand, the maximal possible values of the individual terms $\max_{\mathcal{NS}}\{p_{A_1,A_2}(S_2)\}$ and $\max_{\mathcal{NS}}\{p_{A_1,A_3}(S_2)\}$ decrease all the way from $p(S_3)\geq \frac{3}{4}$. In order to help visualize this, we plot $\max_{\mathcal{NS}}\{p_{A_1,A_2}(S_2)\}$ against $p_{A_1,A_3}(S_2)$ given different observed values of $p(S_3)$ [see FIG. \ref{love}]. For $p(S_3) \in [\frac{3}{4},\frac{5}{6}]$ the maximum sum is still $\frac{3}{2}$, however the individual maximums continually decrease, giving the plots a plateau like shape. 
Again a similar trend is observed under the quantum constraints. We plot $\max_{Q}\{p_{A_1,A_2}(S_2)\}$ against $p_{A_1,A_3}(S_2)$ given different observed values of $p(S_3)$ [see FIG. \ref{love1}]. For $p(S_3) \in [0.68,\frac{3}{4}]$ the maximum sum is still $\frac{3}{2}$, however the individual maximums continually decrease, giving the plots a non-trivial shape.  

\section{Conclusions and future directions}
This work is an attempt to capture a characteristic feature of Nature, namely, \textit{when a system of multiple particles (say spin-$\frac{1}{2}$ particles) manifests genuine non-local correlation, then there cannot be arbitrarily high non-local correlation among any subset of particles}. We refer to this feature, as the \textit{complementarity of genuine multipartite non-locality}. We bring forth this feature without invoking the particulars of the quantum theory. Instead, we show that any no-signaling non-local theory will manifest this property. We use the Svetlichny's criterion for genuine multipartite non-locality and Svetlichny's family of non-local games, to showcase this feature. Svetlichny's non-local games have a recursive structure; i.e., a Svetlichny's non-local game for $n$ parties yields four distinct Svetlichny's non-local games for $k$ parties upon conditioning over inputs and outputs of the $n-k$ parties [see FIG. \ref{prove_me}]. We exploit this feature in-order to prove our main results (\ref{snsn}),(\ref{nvn}) for the no-signaling constraints. Our numerical analysis reveals the fact that these relations tighten under the much stricter quantum constraints [see FIG. \ref{snvssn} and \ref{skvssn}]. \\
However, this feature of no-signaling non-local theories is not restricted to any particular framework. Also, Svetlichny's criterion provides only a sufficient condition for genuine multipartite non-locality. Therefore, for completeness, we provide numerical evidence showcasing this property, by plotting $\max_{\mathcal{NS}}(p(S_2)),\max_{Q}(p(S_2))$ (where $S_2$ is the non-local game equivalent to well known CHSH-expression) against several tripartite measures of non-locality [see \textbf{Appendix A.}]. \\
We compare the notions of \textit{complementarity of genuine multipartite non-locality} and \textit{monogamy of non-locality}. While monogamy of non-locality entails a restriction on possible non-local correlations of an external party with any particular party (or a bunch of parties) within a (genuinely non-locally) 
correlated system, the notion of \textit{complementarity of genuine multipartite non-locality} on the other hand deals with distribution of genuine non-local correlation within a system having some overall genuine correlation [see FIG. \ref{idea}].
While these are distinct features of no-signaling non-local theories, they are subtly related. As a bi-product of this comparison, we find that the monogamy relations tighten with increase in genuine multipartite non-locality for both the no-signaling and the quantum constraints (\ref{monogamy})[see FIG. \ref{monovssn} and \ref{ahCommon}]. \\
Apart from the straightforward relevance to the field of foundations, the complementarity relations presented here have potential applications to the fields of communication complexity, cryptography and game theory. It was shown that for a very large class of multipartite Bell-type inequalities, correlations which violate them lead to an advantage in corresponding communication complexity protocols \cite{brukner2004bell}. And in-turn this advantage in communication complexity protocols is directly related to security of cryptographic protocols \cite{becker1998communication,micciancio2004optimal,guedes2011quantum}. In particular, in protocols involving multiple parties, where the success of the protocol is directly related to amount of violation of genuine multipartite non-locality testing Bell-type inequalities, the complementarity relations provide security against a conspiring (cheating) subgroup of parties. \\
In this work, we deal with how the amount genuine multipartite non-locality of one (arbitrary) subset is restricted due to presence of genuine multipartite non-locality in the system as a whole. It would be interesting to further study the distribution of genuine multipartite non-locality among multiple (disjoint or overlapping) subsets given the presence of genuine multipartite non-locality in the system as a whole.
\section{Acknowledgements}
\noindent AC would like to thank Dr. G Kar, Dr M Banik of PAMU, ISI, Kolkata, India for intuitive advice and D Saha, Dr. M Pawłowski and Dr. K Horodecki of KCIK, Gdansk, Poland for detailed discussions. This project was partly supported by grant, Harmonia 4 (Grant number: UMO-2013/08/M/ST2/00626), ERC AdG QOLAPS.

\bibliography{bibliography}

\clearpage
\input{AppendToMe.tex}
\end{document}

%% file: AppendToMe.tex
\begin{widetext}
\begin{appendices} 
%\subsection{Definition and implications.}

%\section{Appendix A: Tables}
\label{appendix:tables}
Here we provide a brief map of our Appendices. In Appendix A, we provide numerical evidence for our main result, by plotting $\max_{\mathcal{NS}}\{p(S_2)\},\max_{{Q}}\{p(S_2)\}$  against observed values of an assorted bunch of tripartite non-locality measures. \textbf{Appendix B} contains the proof for \textbf{Property \ref{trivially_complementary_recurrence}.}  and \textbf{Property \ref{fixing_one_fixes_another}.}. \textbf{Appendix C}, contains the proof for \textbf{Theorem \ref{thm_1}.}. \textbf{Appendix D}, contains derivation of complementary relations between the winning probability of a $S_2$ game vs. single party marginals and of a $S_n$ game vs. single party marginals and $n-k$ party marginals.  
\textbf{Appendix E}, contains the proof of \textbf{Theorem \ref{thm_2}}.

\section{Other tripartite measures of non-locality vs. CHSH}
In this section, we provide numerical evidence for complementarity between tripartite genuine non-locality and bipartite non-locality using a selected bunch of tripartite measures and $p(S_2)$ as the bipartite measure of non-locality. Let us suppose, that a tripartite non-locality measure, $\mathcal{C}_3$ has an observed value $c$. Now, given this fact, we wish to find $\max_{\mathcal{NS}}p(S_2)$ and approximate $\max_{Q}p(S_2)$. Now, notice that the no-signaling constraints are linear on the conditional probabilities, hence, we solve the linear program of the form,

\begin{equation*}
\begin{aligned}
& {\text{maximize}}
& & p(S_2) \\
& \text{subject to}
& & \mathcal{C}_3=c, \\
&&& p(u_1,u_2,u_3|x_1,x_2,x_3)\in {\mathcal{NS}}.
\end{aligned}
\end{equation*}
Furthermore, we approximate using $Q_2$, a convex set of correlations, which approximates the quantum correlations. Its testing condition is a SDP program in the NPA hierarchy, 
	
\begin{equation*}
\begin{aligned}
& {\text{maximize}}
& & p(S_2) \\
& \text{subject to}
& & \mathcal{C}_3=c, \\
&&& p(u_1,u_2,u_3|x_1,x_2,x_3)\in {Q}_2.
\end{aligned}
\end{equation*}
We use two variations of Svetlichny's non-local game, $M_3,N_3$ \cite{barrett2005nonlocal,pironio2011extremal}. These differ from each-other and from $S_3$ in the way the inputs, $x_1,x_2,x_3$, are included the boolean condition for success. In both cases, complementarity relations are witnessed [see FIG. \ref{MG1},\ref{MG2}]. Next up, we use the Mermin's tripartite expression, $MF_3$. It is a Bell-type (CHSH like) expression, with 
local-real bound, $\max_{LV}\{MF_3\}=2$
equal no-signaling and quantum upper bounds, $\max_{\mathcal{NS}}\{MF_3\}=\max_Q\{MF_3\}=$. As discussed earlier, the violation of Mermin's inequality does not imply genuine multipartite non-locality, but just non-locality. As a result the characteristic complementarity is witnessed in the case of much tighter quantum constraints, \textit{no} complementarity is witnessed at all in the case of no-signaling constraints [see FIG. \ref{MF}]. \\
The correlations which can not be represented as in the form given in (\ref{SL}), are referred to as genuinely tripartite $SV_2$ non-local. While prescribing, (\ref{SL}), Svetlichny tacitly assumed that the measurements can 
 be regarded as simultaneous or the probabilities $p_{{\lambda}_1}(u_1,u_2|x_1x_2 ),p_{\lambda_2}(x_2 x_3|X_2X_3 ),p_{\lambda_3}(u_1,u_3|x_1,x_3 )$ are independent
of the timing of the measurements. This assumption goes well in the case, when the hidden variables are no signaling, otherwise, one might be faced with causality paradoxes. The polytope obtained using no-signaling hidden variables is referred to as $\mathcal{NS}_2$. An alternative solution to the causality puzzle is provided introducing the concept of time ordering in  
(\ref{SL}). The corresponding polytope obtained in this case is called $T_2$. It should be emphasized that $\mathcal{NS}_2 \subset T_2 \subset SV_2$ where the inclusion is strict. We compare $p(S_2)$ with the probabilistic expression, $I_3$. The violation of the corresponding inequality, implies that the correlations are $\mathcal{NS}_2$ and $T_2$ non-local.
Thus, $I_3\le0$ is a weaker (less stricter) inequality compared to the Svetlichny's inequality \cite{bancal2011definition}. Here, the complementarity is witnessed all the way from $I_3>0$ under both no-signaling and quantum constraints [see FIG. \ref{I}]. \\
Finally, we use the guess your neighbor's input game ($GYNI_n$) \cite{acin2016guess,pironio2011extremal}. This game consists of $n$ players, arranged on a ring, wherein each player receives a binary input $x_i$ for $i\in \{1,\ldots n\}$. The aim of the game is, to obtain a situation, wherein each player outputs a bit $u_i$ equal to the input bit of it's right neighbor, $u_i=x_{i-1}$. In case of $3$ players, the game is denoted as, $GYNI_{3}$. Here again, one can clearly visualize the complementarity relations [see FIG. \ref{GYNI}]. The expressions for these measures, their respective quantum and no-signaling bounds are presented in the Table \ref{bounds}. The plots for complementarity relations can be found in FIG. \ref{myTableSucks}. 
\begin{table}[!htb]
\centering
\begin{tabular} {|c|c|c|c|c|} 
\hline
%\diag{.1em}{3.48cm}{}{}
%{$p_{win}^{S_{n-1}}$}{$u_n,x_n$}
Tripartite Measure  \; &Expression \; & Quantum bound \;& No-signaling bound\; \\ \hline
$M_3$ & $u_1\oplus u_2\oplus u_3= x_1x_2x_3$ & $\frac{7}{8}$ & 1\\[5pt] \hline
$N_3$ & $u_1\oplus u_2\oplus u_3= x_1x_2 \oplus x_2x_3$&$\approx 0.7818$ & 1\\[5pt] \hline
$MF_3$ & \shortstack{~ $\langle x_1=1,x_2=0,x_3=0\rangle +\langle x_1=0,x_2=1,x_3=0\rangle$ \\ $+\langle x_1=0,x_2=0,x_3=1\rangle -\langle x_1=0,x_2=1,x_3=1\rangle$}& 4  & 4\\[5pt] \hline
$I_3$& \shortstack{~$-2p_{A_1A_2}(u_1=0,u_2=0|x_1=1,x_2=1)$ \\$-2p_{A_2A_3}(u_2=0,u_3=0|x_2=1,x_3=1)$ \\$-2p_{A_3A_1}(u_3=0,u_1=0|x_3=1,x_1=1)$\\$-p(u_1=0,u_2=0,u_3=0|x_1=0,x_2=0,x_3=1)$\\$-p(u_1=0,u_2=0,u_3=0|x_1=0,x_2=1,x_3=0)$\\$-p(u_1=0,u_2=0,u_3=0|x_1=1,x_2=0,x_3=0)$\\$+2p(u_1=0,u_2=0,u_3=0|x_1=0,x_2=1,x_3=1)$\\$+2p(u_1=0,u_2=0,u_3=0|x_1=1,x_2=0,x_3=1)$\\$+2p(u_1=0,u_2=0,u_3=0|x_1=1,x_2=1,x_3=0)$\\$+2p(u_1=0,u_2=0,u_3=0|x_1=1,x_2=1,x_3=1)$}& 0.14 & 0.5\\[5pt] \hline
$p(GYNI_3)$ & \shortstack{~ $p(u_1=0,u_2=0,u_3=0|x_1=0,x_2=0,x_3=0)$\\$+p(u_1=1,u_2=1,u_3=0|x_1=0,x_2=1,x_3=1)$\\$+p(u_1=0,u_2=1,u_3=1|x_1=1,x_2=0,x_3=1)$\\$+p(u_1=1,u_2=0,u_3=1|x_1=1,x_2=1,x_3=0)$}& 0.25 & $\frac{1}{3}$\\[5pt] \hline
\end{tabular}
\caption{Expressions for tripartite measures of non-locality, and their corresponding quantum and no-signaling upper bounds.} 
\label{Sncase2}
\label{bounds}
\end{table}

\begin{figure*}
\subfloat[\label{MG1} The observed winning probability in the $M_3$ game, $p(M_3)$.]{%
  \includegraphics[scale=0.6]{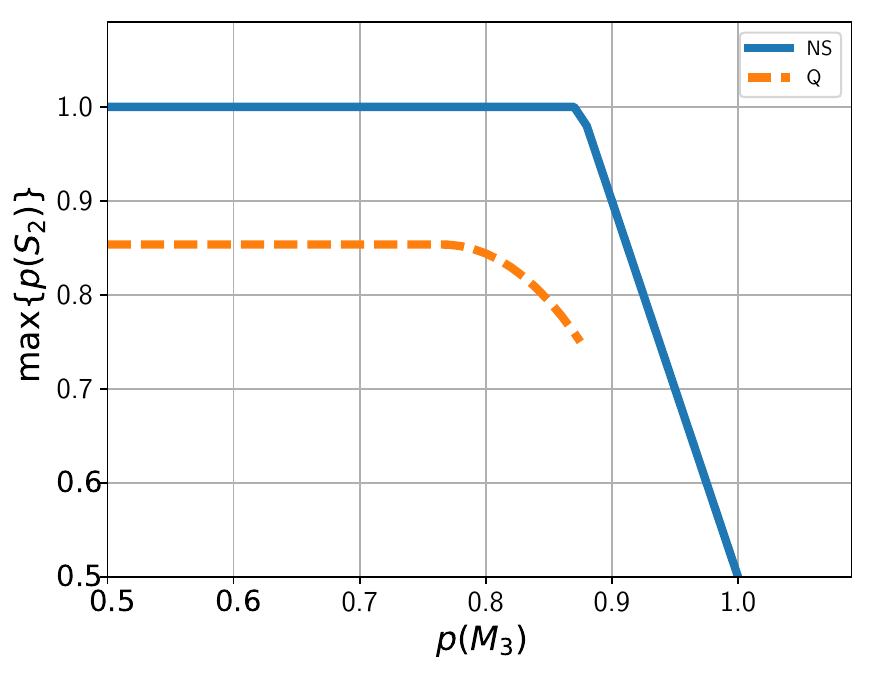}%
} \hfill
\subfloat[\label{MG2} The observed winning probability in the $N_3$ game, $p(N_3)$.]{%
  \includegraphics[scale=0.6]{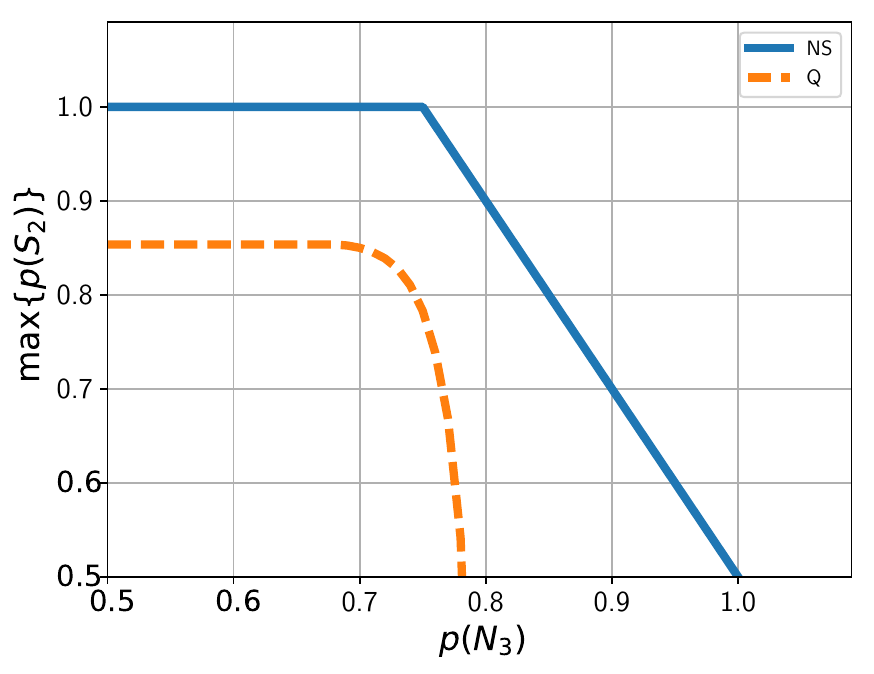}%
} \hfill
\subfloat[\label{MF} The observed value of Mermin's facet expression, $MF_3$.]{%
  \includegraphics[scale=0.6]{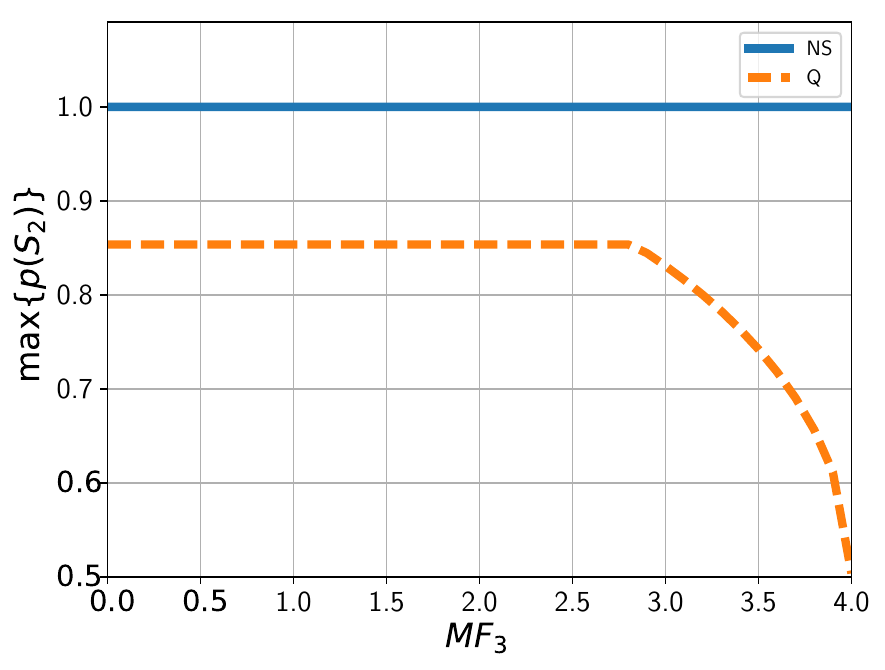}%
} \hfill
\subfloat[\label{I} The observed value of $I_3$.]{%
  \includegraphics[scale=0.6]{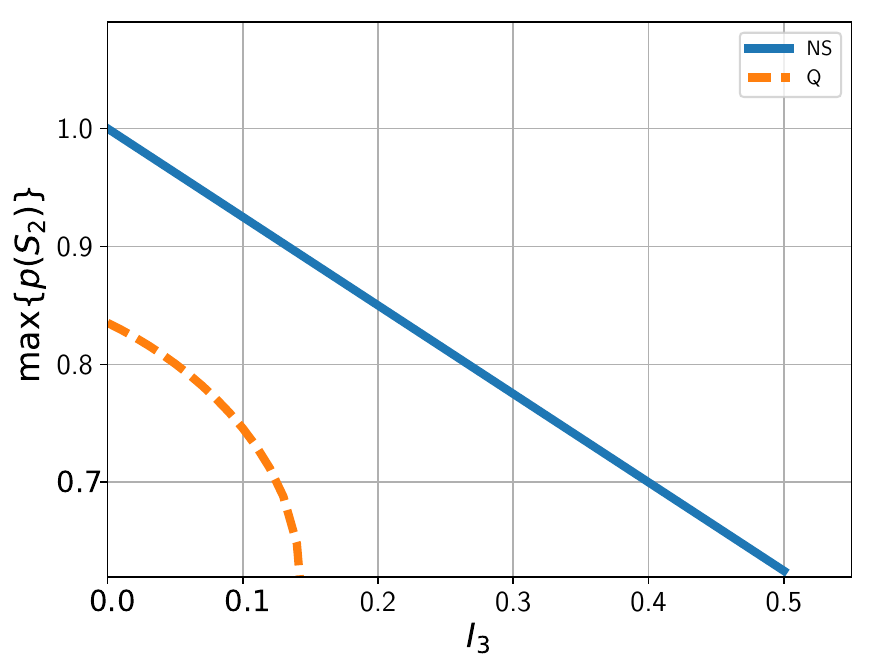}%
} \hfill
\subfloat[\label{GYNI} The observed winning probability in the GYNI game, $p(GYNI_3)$.]{%
  \includegraphics[scale=0.6]{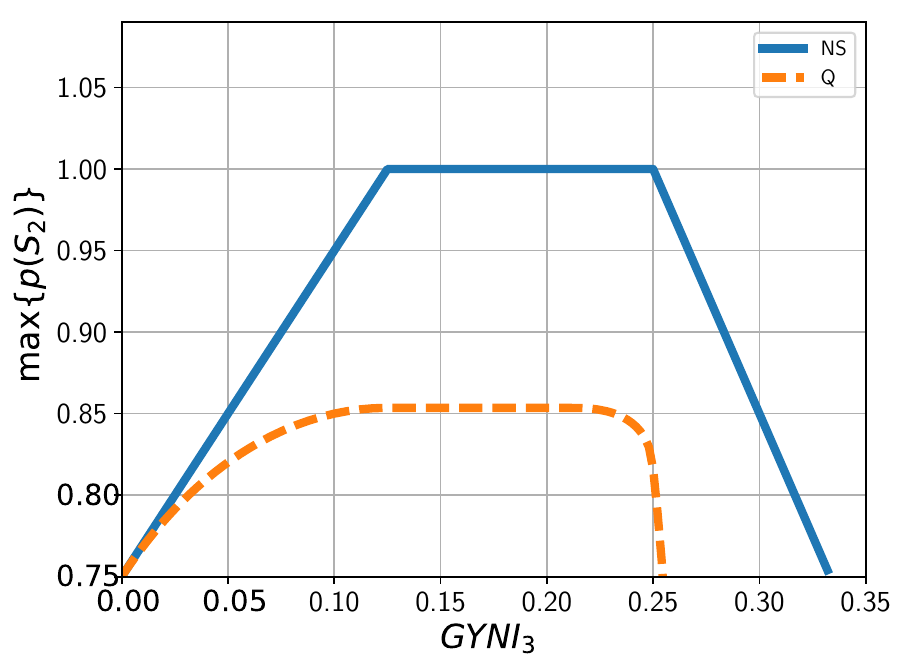}%
} \hfill
\caption{Plot of maximum achievable winning probability in a $S_2$ game, $\max \{p(S_2)\}$ against some tripartite measures of non-locality. For maximization over no-signaling probability distributions, the solid blue line, and for quantum probabilities, the dashed orange curve.
}
\label{myTableSucks}
\end{figure*}

\section{Decomposition of $S_n$-games}
\label{svet_prop}
\noindent A $n$-party Svetlichny's non-local game, $S_n^i$, reduces to four $n-1$-party Svetlichny's non-local games, given the fixed assignment of the input and output bits of the $n$th party, denoted by $S_{n-1}^j\equiv S_n^i|(u_n=u_n',x_n=x_n')$, for  $u_n',x_n' \in \{0,1\}$. Similarly, a $S_n^j$ game reduces to four $k$-party Svetlichny's non-local games, given the fixed assignment of the input and output bits of $n-k$ parties, denoted by $S_{k}^j\equiv S_n^i|(u_{k+1}=u_{k+1}',\ldots u_n=u_n',x_{k+1}=x_{k+1}',\ldots x_n=x_n')$ where $u_i',x_i'\in \{0,1\}$ for $k < i \le n$. If $S_n^i$ game is of the form (\ref{svetlichny_recurssive}), then the $S_{k}^j$ game, thus obtained, is given by,
%\begin{widetext}
\begin{eqnarray}
\label{svetlichny_in_any_lower_form}
&&(\bigoplus_{1\le i \le k}u_i) \oplus (\bigoplus_{k < i \le n}u_i{'}) \oplus c_0 ={}\nonumber\\&&
(\bigoplus_{1\le i<j \le k}x_i x_j) \oplus
 (\bigoplus_{1 \le i \le k} (\bigoplus_{k<j\le n}x_{j}{'} \oplus c_i ) \ x_i) \oplus
 (\bigoplus_{k<i \le n}c_ix_i{'}) \; . {}\nonumber\\&&  
\end{eqnarray} 
Using the definitions of $\mathcal{A}_{n-k},\mathcal{B}_{n-k}$, we can rewrite the above as,
%\begin{widetext}
\begin{eqnarray}
\label{svetlichny_k_00}
&& S_k^1 \equiv S_{n}^j|( \mathcal{A}_{n-k}=0,\mathcal{B}_{n-k}=0) \equiv  {}\nonumber\\&&
\bigoplus_{1 \le i\le k} u_i \oplus c_0 = 
(\bigoplus_{1 \le i<j \le k} x_i x_j) \oplus (\bigoplus_{1 \le i\le k}c_i x_i) {} \; ,\nonumber\\&& 
{S_k^1}' \equiv S_{n}^j|(\mathcal{A}_{n-k}=1,\mathcal{B}_{n-k}=0) \equiv {}\nonumber\\&&  
\bigoplus_{1 \le i\le k} u_i \oplus c_0 \oplus 1 = 
(\bigoplus_{1 \le i<j \le k} x_i x_j) \oplus (\bigoplus_{1 \le i\le k}c_i x_i) {} \ ,\nonumber\\&&  
S_k^2 \equiv S_{n}^j|(\mathcal{A}_{n-k}=0,\mathcal{B}_{n-k}=1) \equiv {}\nonumber\\&&  
\bigoplus_{1\le i\le k} u_i \oplus c_0 =
 (\bigoplus_{1 \le i<j \le k} x_i x_j) \oplus (\bigoplus_{1 \le i\le k}(1\oplus c_i) x_i) {} \ ,\nonumber\\&& 
{S_k^2}'\equiv S_{n}^j|(\mathcal{A}_{n-k}=1,\mathcal{B}_{n-k}=1) \equiv {}\nonumber\\&&
\bigoplus_{1 \le i \le k} u_i \oplus c_0 \oplus 1 = 
(\bigoplus_{1 \le i,j \le k} x_i x_j) \oplus  (\bigoplus_{1 \le i \le k}(1\oplus c_i) x_i)  {} \; .\nonumber\\&& 
\end{eqnarray}
The equation (\ref{svetlichny_k_00}) reveals that the game $S_n^j$ game decomposes into one of the four games: ${S_k^1},{S_k^1}',{S_k^2},{S_k^2}'$ depending on the value of the boolean expressions $\mathcal{A}_{n-k}$ and $\mathcal{B}_{n-k}$. Moreover, the following relation holds between the $S_k^1$ game and $S_k^2$ game,
\begin{eqnarray}
\label{rel_for_fiu_one_fiu_another}
&&S_k^1  \equiv
S_k^2 \oplus \bigoplus_{1 \le i \le k}x_i \; .
\end{eqnarray}
Also from \textbf{Property \ref{trivially_complementary_recurrence}.}, it follows that the same relation holds between ${S_k^1}'$ game and ${S_k^2}'$ game, respectively. \\   

%\end{widetext}

\section{ Proof of Theorem 1}
We shall now give the proof for \textbf{Theorem \ref{thm_1}.} using induction. First, we show that the relations hold for $n=1$. 
\subsection*{$S_1$ vs. $S_1$}
Consider the four single party games $S_1^1\equiv u_1=0$, ${S_1^1}'\equiv u_1=1$, ${S_1^2}\equiv u_1=x_1$ and ${S_1^2}'\equiv x_1\oplus 1$  obtained from (\ref{svetlichny_recurssive}) for $n=1$. We want to find, $\max\{p(S_1^2)\}$ for a given value of $p({S_1^1}) \in [\frac{1}{2},1] $. Using the fact that, $p({S_1^1})=\frac{p(u_1=0|x_1=0)+p(u_1=0|x_1=1)}{2}$ and $p({S_1^2})=\frac{p(u_1=0|x_1=0)+p(u_1=1|x_1=1)}{2}$, we get $p({S_1^2}) + p({S_1^1}) = \frac{1+2p(u_1=0|x_1=0)}{2}$. Now, $\max\{p(u_1=0|x_1=0)\}=1$, which leads one to $\max\{p(S_1^2)\}=\frac{3}{2}-p({S_1^1})$. Notice, for $p({S_1^1})\in [0,\frac{1}{2})$, $p(u_1=0|x_1=0)$ cannot be $1$. So we shall consider, $p({S_1^2}) - p({S_1^1}) = \frac{1-2p(u_1=0|x_1=1)}{2}$. Now, $\min\{p(u_1=0|x_1=1)\}=0$, which yields, $\max\{p(S_1^2)\}=\frac{1}{2}+p({S_1^1})=\frac{3}{2}-p({S_1^1}')$.  
So in general, we have for a given value $p(S_1^i)$, $\max\{p(S_1^j)\}$ is given by,
\begin{eqnarray}
\label{marginal_game}
    \max_{\mathcal{NS}} \{ p({S_1^j}) \} =
    \begin{cases}
      p({S_1^i}), & \text{if}\ S_1^j \equiv S_1^i \\
      1 - p({S_1^i}), & \text{if}\ S_1^j \equiv {S_1^i}{'}   \\
      p({S_1^i})+\frac{1}{2}, &  \text{if} \ S_1^j \not \equiv \{{S_1^i},{S_1^i}{'}\},  p({S_1^i})\le \frac{1}{2} \\
      
      -p({S_1^i}) +\frac{3}{2}, & \text{if} \ S_1^j \not \equiv \{{S_1^i},{S_1^i}{'}\},  p({S_1^i}) \ge \frac{1}{2} \ . \\
     
    \end{cases} 
\end{eqnarray}
\noindent
Notice that here we \textit{did not} use any constrains on the conditional probability distribution $p(u_1|x_1)$, hence, this stands true for classical, quantum and no-signaling restrictions.\\
\subsection*{$S_n$ vs. $S_n$}
Now for the proof for any $n$, as the next step of induction, we assume that the complementary relation presented in \textbf{Theorem 1.}, holds for $S_{n-1}$ games, and consequently prove 
the same for $S_n$ games. Let us consider two $S_n$ games, $S_n^i$ and $S_n^j$ which are neither equivalent nor trivially 
complementary to each other. In addition, we consider only the case when $ p(S_n^i)\geq \frac{1}{2}$. The winning probability of the $S_n^i$ game can be expressed as,
\begin{eqnarray}
\label{S_n__vs__S_n__1}
%\begin{eqnaraay}
&&p({S_n^i}) ={}\nonumber\\&& \frac{1}{2} \bigg( p(u_n=0|x_n=0)(p({S_{n-1}^1}|u_n=0,x_n=0){}\nonumber\\&&+p(u_n=1|x_n=0)(p({{S_{n-1}^1}{'}}|u_n=1,x_n=0){}\nonumber\\&&
+p(u_n=0|x_n=1)(p({S_{n-1}^2}|u_n=0,x_n=1){}\nonumber\\&&+p(u_n=1|x_n=1)(p({{S_{n-1}^2}{'}}|u_n=1,x_n=1) \bigg) \; ,
\nonumber\\&&
%\end{split} 
\end{eqnarray}
where ${S_{n-1}^1} \equiv S_{n}^i||(u_n=0,x_n=0)$,
${{S_{n-1}^1}{'}} \equiv S_{n}^i|(u_n=1,x_n=0)$,${S_{n-1}^2} \equiv S_{n}^i||(u_n=0,x_n=1)$,
${{S_{n-1}^2}{'}} \equiv S_{n}^i|(u_n=1,x_n=1)$.  Similarly, the winning probability of the $S_n^i$ game can be expressed as, 
\begin{eqnarray}
\label{S_n__vs__S_n__2}
%\begin{eqnaraay}
&&p({S_n^j}) ={}\nonumber\\&& \frac{1}{2} \bigg( p(u_n=0|x_n=0)(p({S_{n-1}^3}|u_n=0,x_n=0){}\nonumber\\&&+p(u_n=1|x_n=0)(p({S_{n-1}^3}{'}|u_n=1,x_n=0){}\nonumber\\&&
+p(u_n=0|x_n=1)(p({S_{n-1}^4}|u_n=0,x_n=1){}\nonumber\\&&+p(u_n=1|x_n=1)(p({{S_{n-1}^4}{'}}|u_n=1,x_n=1) \bigg) \; . \nonumber\\&& 
%\end{split}
\end{eqnarray}
Now, one can write no-signaling condition as,
\begin{equation}
\begin{split}
p(u_n=0|x_n=0)p(u_1,\ldots u_{n-1}|x_1,\ldots x_n)_{u_n=0,x_n=0}+p(u_n=1|x_n=0)p(u_1,\ldots u_{n-1}|x_1,\ldots x_n)_{u_n=1,x_n=0}= \\
p(u_n=0|x_n=1)p(u_1,\ldots u_{n-1}|x_1,\ldots x_{n-1})_{u_n=1,x_n=1}+p(u_n=1|x_n=1)p(u_1,\ldots u_{n-1}|x_1,\ldots x_{n-1})_{u_n=1,x_n=1}.
\end{split}
\end{equation}
Notice that the winning probability of any $S_{n-1}$ game is sum over specific probabilities of the form $p(u_1,u_2,\ldots,u_{n-1}|x_1,x_2,\ldots,x_{n-1})$, therefore we can state the no-signaling condition for any particular $S_{n-1}^j$ game,
\begin{equation}
\label{no_signaling_conx_indp_of_input}
\begin{split}
&~~p(u_n=0|x_n=0)(p(S_{n-1}^j|u_n=0,x_n=0))  \\ &+p(u_n=1|x_n=0)(p(S_{n-1}^j|u_n=1,x_n=0))  \\
&= p(u_n=0|x_n=1)(p(S_{n-1}^j|u_n=0,x_n=1))  \\ &+p(u_n=1|x_n=1)(p(S_{n-1}^j|u_n=1,x_n=1))\; .
\end{split}
\end{equation}
% In addition to Eq. \ref{S_n__no_signaling}, no-signaling condition necessitates that values of $p_{win}^{s_{n-1}}$ for $S_{n-1}$ games are in accordance with Eq. \ref{marginal_game} for any fixed values of $u_n$ and $x_n$. \\
\noindent From \textbf{Property \ref{trivially_complementary_recurrence}.} and \textbf{Property \ref{fixing_one_fixes_another}.} we have the following two representative cases:
\begin{enumerate}
\item $ S_{n-1}^1,S_{n-1}^2  \not \in \{S_{n-1}^3, {S_{n-1}^3}{'},S_{n-1}^4, {S_n^4}{'}\}$: Following from the observation in ($S_n$ vs. marginals), we shall fix the marginals of the $n$th party to be $\frac{1}{2}$, i.e., $p(u_n|x_n)=\frac{1}{2}$. Now, for $p(S_n^i) \ge \frac{1}{2}$ let us suppose that, $p(S_n^i)=\frac{1}{2}+\delta$ for some $\delta\in [0,\frac{1}{2}]$. Furthermore, we assume w.l.o.g. that, $p(S_{n-1}^1|u_n=0,x_n=0)=\frac{1}{2}+\delta_1$, $(p({S_{n-1}^1}{'}|u_n=1,x_n=0)=\frac{1}{2}+\delta_2$,  $p(S_{n-1}^2|u_n=0,x_n=1)=\frac{1}{2}+\delta_3$ and  $p({S_{n-1}^2}{'}|u_n=1,x_n=1)=\frac{1}{2}+\delta_4$ where $\delta_i \in [-\frac{1}{2},\frac{1}{2}]$ for $i \in \{0,1,2,3\}$. Therefore, $\delta=\frac{1}{4}\sum_i\delta_i$.
Now, from the assumption of involved in induction, that the $S_{n-1}$ follow the relation presented in \textbf{Theorem \ref{thm_1}.}, we obtain $\max_{\mathcal{NS}} \{p(S_n^j)\}=1-\frac{1}{4}\sum_i\delta_i=1-\delta=\frac{3}{2}-p({S_n^i})$. For the purpose of illustration, we provide the optimal assignment in the form of the Table \ref{Sncase1}, with $\delta_1=\delta_2=\delta_3=\delta_4=\delta$. Notice, that sum of values in first two columns is equal to that of the last two columns, thus condition in (\ref{no_signaling_conx_indp_of_input}) is satisfied. Also, all $S_{n-1}$ games satisfy the complementary relations given in \textbf{Theorem \ref{thm_1}.}.
\begin{table}[!htb]
\centering
\begin{tabular} {|l|l|l|l|l|} 
\hline
%\diag{.1em}{3.48cm}{}{}
%{$p_{win}^{S_{n-1}}$}{$u_n,x_n$}
&$u_n=0,x_n=0$&$u_n=1,x_n=0$&$u_n=0,x_n=1$&$u_n=1,x_n=1$\tabularnewline \hline
$p({S_{n-1}^1}|u_n,x_n)$& $\frac{1}{2}+\delta$ & $\frac{1}{2}-\delta$ & $1-\delta$&$\delta$\tabularnewline \hline
$p({S_{n-1}^1}{'}|u_n,x_n)$& $\frac{1}{2}-\delta$ & $\frac{1}{2}+\delta$ & $\delta$ &$1-\delta$\tabularnewline \hline
$p({S_{n-1}^2}|u_n,x_n)$&$1-\delta$ & $\delta$ & $\frac{1}{2}+\delta$&$\frac{1}{2}-\delta$\tabularnewline \hline
$p({S_{n-1}^2}{'}|u_n,x_n)$& $\delta$ & $1-\delta$ & $\frac{1}{2}-\delta$ &$\frac{1}{2}+\delta$\tabularnewline \hline
\end{tabular}
\caption{\label{Sncase1} Optimal assignment for case 1. }
\end{table}
\item $S_{n-1}^1 \equiv S_{n-1}^3$   and $S_{n-1}^2  \equiv {S_{n-1}^4}{'}$: Here, proceeding with notations and assumptions of the previous case, we find that it is possible to fix $(p({S_{n-1}^1}|u_n=0,x_n=0)=(p({{S_{n-1}^1}{'}}|u_n=1,x_n=0)=1$. Such that $\delta=\frac{1+\delta_3+\delta_4}{4}$. Again, from the assumption that $S_{n-1}$ games follow the relation presented in \textbf{Theorem \ref{thm_1}.}, we obtain $\max_{\mathcal{NS}}\{p(S_n^j)\}=1-\frac{(1+\delta_3+\delta_4)}{4}=1-\delta=\frac{3}{2}-p(S_n^i)$. Yet again, for the purpose of illustration, we provide the optimal assignment in the form of the Table \ref{Sncase1}, with $\delta_1=\delta_2=\delta_3=\delta_4=\delta$. Notice, that sum of values in first two columns is equal to that of the last two columns, thus condition in (\ref{no_signaling_conx_indp_of_input}) is satisfied. Also, all $S_{n-1}$ games satisfy the complementary relations given in \textbf{Theorem \ref{thm_1}.}.
\begin{table}[!htb]
\centering
\begin{tabular} {|l|l|l|l|l|} 
\hline
%\diag{.1em}{3.48cm}{}{}
%{$p_{win}^{S_{n-1}}$}{$u_n,x_n$}
&$u_n{'}=0,x_n{'}=0$&$u_n{'}=1,x_n{'}=0$&$u_n{'}=0,x_n{'}=1$&$u_n{'}=1,x_n{'}=1$\\ \hline

$p({S_{n-1}^1}'|u_n,x_n)$& $1$ & $0$ & $\frac{1}{2}$&$\frac{1}{2}$\\ \hline

$p({S_{n-1}^1}'|u_n,x_n)$& $0$ & $1$ & $\frac{1}{2}$ &$\frac{1}{2}$\\ \hline

$p({S_{n-1}^2}|u_n,x_n)$&$\frac{1}{2}$ & $\frac{1}{2}$ & $2\delta$&$1-2\delta$\\ \hline

$p({S_{n-1}^2}'|u_n,x_n)$& $\frac{1}{2}$ & $\frac{1}{2}$ & $1-2\delta$ &$2\delta$\\ \hline
\end{tabular}

\caption{Optimal assignment for case 2.} 
\label{Sncase2}
\end{table}
\end{enumerate} 
All other cases, can be handled using the above to cases.
\section{Complementarity with Marginals}
\subsection*{ $S_2$ {vs. marginals} }
\noindent
Consider a bipartite Svetlichny's non-local game, $S_2 \equiv u_1 \oplus u_2 =x_1x_2$. Its winning probability can be rewritten as,
%\begin{widetext}
\begin{eqnarray}
\label{Chsh_vs_marginal_pwin_breakdown}
%\begin{eqnaraay}
p(S_2) &&= \frac{1}{4} \bigg( p(u_2=0|x_2=0)(p(u_1=0|x_1=0)+p(u_1=0|x_1=1))_{u_2=0,x_2=0}
{}\nonumber\\&&+p(u_2=1|x_2=0)(p(u_1=1|x_1=0)+p(u_1=1|x_1=1))_{u_2=1,x_2=0}
{}\nonumber\\&&
+p(u_2=0|x_2=1)(p(u_1=0|x_1=0)+p(u_1=1|x_1=1))_{u_2=0,x_2=1}{}\nonumber\\&&+p(u_2=1|x_2=1)(p(u_1=1|x_1=0)+p(u_1=0|x_1=1))_{u_2=1,x_2=1} \bigg) {} \; ,
%\end{split} 
\end{eqnarray}
%\end{widetext}
%\begin{equation}
%p_{win}^{CHSH}=\frac{\sum_{X.Y=a\uplus b}p(a,b|X,Y)}{4}
%\end{equation}
%which may also is written in the form,
%\begin{equation}\label{CHSHtut}
%p^{CHSH}_{win}=\sum_{i,j}{\frac{p(b=i|Y=j)\sum_{X.j=a\oplus i}(a |X,b=i,Y=j)}{4}}
%\end{equation}
\noindent
where $()_{u_2=i,x_2=j}$ indicates the fact that all probabilities contained within the brackets are conditioned on $u_2=i,x_2=j$. Let $p(u_2=0|x_2=0)=\mathsf{p}$ and $p(u_2=0|x_2=1)=\mathsf{p}'$, 
then no-signaling condition
from $A_2$ to $A_1$ can be stated as,
%\begin{widetext}
\begin{eqnarray}
\label{chsh_vs_marginal_no_signaling_condition_1}
&&\mathsf{p}p(u_1=0|x_1=0)_{u_2=0,x_2=0}+(1-\mathsf{p})p(u_1=0|x_1=0)_{u_2=1,x_2=0}={}\nonumber\\&&
\mathsf{p}'p(u_1=0|x_1=0)_{u_2=0,x_2=1}+(1-\mathsf{p}')p(u_1=0|x_1=0)_{u_2=1,x_2=1} \; ,
\end{eqnarray}
\begin{eqnarray}
\label{chsh_vs_marginal_no_signaling_condition_2}
&&\mathsf{p}p(u_1=0|x_1=1)_{u_2=0,x_2=0}+(1-\mathsf{p})p(u_1=0|x_1=1)_{u_2=1,x_2=0}={}\nonumber\\&&
\mathsf{p}'p(u_1=0|x_1=1)_{u_2=0,x_2=1}+(1-\mathsf{p}')p(u_1=0|x_1=1)_{u_2=1,x_2=1} \; .
\end{eqnarray}
%\end{widetext}
Now, in-order to find $\max_{\mathcal{NS}} \{p({S_2})\}$, we fix
$p(u_1=0|x_1=0)_{u_2=0,x_2=0}=1$ ,$p(u_1=1|x_1=0)_{u_2=1,x_2=0}=1$,$p(u_1=0|x_1=0)_{u_2=0,x_2=1}=1$ and $p(u_1=1|x_2=0)_{u_2=1,x_2=1}=1$ and obtain $\mathsf{p}=\mathsf{p}{'}$ from (\ref{chsh_vs_marginal_no_signaling_condition_1}). Further fixing $p(u_1=0|x_1=1)_{u_2=0,x_2=0}=1$,  $p(u_1=1|x_1=1)_{u_2=1,x_2=0}=1$,  $p(u_1=1|x_1=1)_{u_2=0,x_2=1}=1$, we obtain $p(u_1=0|x_1=1)_{u_2=1,x_2=1}=\frac{\mathsf{p}}{1-\mathsf{p}}$ for $p\in [0,\frac{1}{2}] $ and  $p(u_1=0|x_2=1)_{u_2=1,x_2=1}=\frac{1-\mathsf{p}}{\mathsf{p}}$ for $\mathsf{p}\in [\frac{1}{2},1] $ from  (\ref{chsh_vs_marginal_no_signaling_condition_2}). This in-turn leads us to, 
\begin{equation}
\label{chsh_marginal_result}
    \max_{\mathcal{NS}} \{ p({S_2})\} =
    \begin{cases}
      \frac{3+2\mathsf{p}}{4}, & \text{if}\ \mathsf{p}\in[0,\frac{1}{2}]\\
      \frac{5-2\mathsf{p}}{4}, & \text{if}\ \mathsf{p}\in[\frac{1}{2},1] \; .\\
     \end{cases} 
\end{equation}

\subsection*{${S_n}$ {vs. marginals}}
\noindent
Now we prove that the complementary relation of the type given by ( \ref{chsh_marginal_result}), hold for $S_n$ games with $n \ge 3$. W.l.o.g., we consider, $S_n \equiv  \bigoplus_{1\le i \le n} u_i = \bigoplus_{1\le i<j \le n} x_i x_j$. We define $A=\bigoplus_{1\le i \le n-1}u_i$, $B=\bigoplus_{1\le i<j \le n-1}x_ix_j$ and $C=\bigoplus_{1\le i \le n-1}x_i$. Also let $p(u_n=0|x_n=0)=\mathsf{p}$ and $p(u_n=0|x_n=1)=\mathsf{p}{'}$. Then the winning probability of the $S_n$-game can be rewritten as,
\begin{eqnarray}
\label{S_n_vs_marginal_win_probability}
p({S_n^i}) &&= \frac{1}{2^n} \Bigg( p(u_n=0|x_n=0)(p(A=B))_{u_n=0,x_n=0}
{}\nonumber\\&&+p(u_n=1|x_n=0)(p(A=B\oplus1))_{u_n=1,x_n=0}{}\nonumber\\&&
+p(u_n=0|x_n=1)(p(A=B\oplus C))_{u_n=0,x_n=1}
{}\nonumber\\&&+p(u_n=1|x_n=1)(p(A=B\oplus C \oplus 1))_{u_n=1,x_n=1} \Bigg)  \; ,
\end{eqnarray}
where $()_{u_2=i,x_2=j}$ indicates the fact that all probabilities contained within the brackets are conditioned on $u_2=i,x_2=j$.To proceed further, we consider the following two cases:
\begin{enumerate}
\item First we consider an assignment of input bits, such that $C=0$. The no-signaling can be rewritten as,
\begin{eqnarray}
\label{S_n_vs_marginal_case_1}
\mathsf{p}(p(A=B))_{u_n=0,x_n=0}+(1-\mathsf{p})(p(A=B))_{u_n=1,x_n=0} \\=
\mathsf{p}{'}(p(A=B))_{u_n=0,x_n=1}+(1-\mathsf{p}{'})(p(A=B))_{u_n=1,x_n=1}
{}\nonumber \; .
\end{eqnarray}
To maximize $p({S_n})$ we keep $(p(A=B))_{u_n=0,x_n=0}=(p(A=B))_{u_n=0,x_n=1}=1$ and $(p(A=B))_{u_n=1,x_n=0}=(p(A=B))_{u_n=1,x_n=1}=0$ in (\ref{S_n_vs_marginal_case_1}), obtaining $\mathsf{p}=\mathsf{p}{'}$.
\\
\item Now, we consider an assignment of input bits, such that $C=1$. In this case, in-order to maximize $p({S_n})$, we keep 
$(p(A=B))_{u_n=0,x_n=0}=1$ and $(p(A=B))_{u_n=1,x_n=0}=(p(A=B))_{u_n=0,x_n=1}=0$ in (\ref{S_n_vs_marginal_case_1}), getting $(p(A=B))_{u_n=1,x_n=1}=\frac{\mathsf{p}}{1-\mathsf{p}}$ for $\mathsf{p} \le \frac{1}{2}$ (and $\frac{1-\mathsf{p}}{\mathsf{p}}$ for $\mathsf{p} \ge \frac{1}{2}$, obtained by replacing $\mathsf{p}$ by $1-\mathsf{p}$).
\end{enumerate}
As there are $2^{n-2}$ assignment of input bits possible, for $C=0$ and $C=1$, keeping the values of terms (\ref{S_n_vs_marginal_win_probability}) according to that described in above two cases, we get the same expression for $\max_{\mathcal{NS}}  \{p({S_n})\}$, as given in (\ref{chsh_marginal_result}).

\subsection*{${S_n}$ {vs. auxiliary games}}
We again consider the following $S_n$ game $S_n^i \equiv  \bigoplus_{1\le i \le n} u_i = \bigoplus_{1\le i<j \le n} x_i x_j$ ( it can be proved similarly for other $S_n$ games ).
We define $A=\bigoplus_{1\le i \le k}u_i$, $B=\bigoplus_{1\le i<j \le k}x_ix_j$ and $C=\bigoplus_{1\le i \le k}x_i$. Also fix $p(\mathcal{A}_{n-k}=0|\mathcal{B}_{n-k}=0)=p$ and $p(\mathcal{A}_{n-k}=0|\mathcal{B}_{n-k}=1)=p{'}$ ( where $A_n^k$ and $B_n^k$ are defined according to Eq. \ref{aux1} and \ref{aux2} ). Using Eq. \ref{general_win_probability2}

\begin{eqnarray}
\label{S_n_vs_marginal_win_probability2}
p_{win}^{S_n^i} &&= \frac{1}{2^{k+1}} \Bigg( p(\mathcal{A}_{n-k}=0|\mathcal{B}_{n-k}=0)\bigg(\sum_{x_1{'},\ldots ,x_{k}{'} \in \{0,1\}}p(A=B|x_1=x_1{'},\ldots x_{k}=x_{k}{'})\bigg)_{\mathcal{A}_{n-k}=0,\mathcal{B}_{n-k}=0}
{}\nonumber\\&&+p(\mathcal{A}_{n-k}=1|\mathcal{B}_{n-k}=0)\bigg(\sum_{x_1{'},\ldots x_{k}{'} \in \{0,1\}}p(A=B\oplus1|x_1=x_1{'},\ldots x_{k}=x_{k}{'})\bigg)_{\mathcal{A}_{n-k}=1,\mathcal{B}_{n-k}=0}{}\nonumber\\&&
+p(\mathcal{A}_{n-k}=0|\mathcal{B}_{n-k}=1)\bigg(\sum_{x_1{'},\ldots x_{k}{'} \in \{0,1\}}p(A=B\oplus C|x_1=x_1{'},\ldots x_{k}=x_{k}{'})\bigg)_{\mathcal{A}_{n-k}=0,\mathcal{B}_{n-k}=1}
{}\nonumber\\&&+p(\mathcal{A}_{n-k}=1|\mathcal{B}_{n-k}=1)\bigg(\sum_{x_1{'},\ldots x_{k}{'} \in \{0,1\}}p(A=B\oplus C \oplus 1|x_1=x_1{'},\ldots x_{k}=x_{k}{'})\bigg)_{\mathcal{A}_{n-k}=1,\mathcal{B}_{n-k}=1} \Bigg)  \; ,
\end{eqnarray}
where $()_{\mathcal{A}_{n-k}=i,\mathcal{B}_{n-k}=j}$ indicates the fact that all probabilities contained within the brackets are conditioned on $\mathcal{A}_{n-k}=i,\mathcal{B}_{n-k}=j$.
From this point onwards, following the same steps as above, one obtains the same expression for $\max_{\mathcal{NS}}  \{p({S_n})\}$, as given in (\ref{chsh_marginal_result}).

\section {Proof of Theorem 2}
Here, we prove the relation presented in \textbf{Theorem 2.}. First, we find the value of $\max_{\mathcal{NS}}\{p({S_{n-1}^j})\}$ for a given value of $p({S_n^i})\ge \frac{1}{2}$.
\subsection*{$S_{n-1}$ vs. $S_n$}
The winning probability of the $S_{n}^i$ game can be re-written as,
\begin{eqnarray}
\label{temp}
%\begin{eqnaraay}
p({S_n^i}) &&= \frac{1}{2} \bigg( p(u_n=0|x_n=0)(p(S_{n-1}^1|{u_n=0,x_n=0}){}\nonumber\\&&+p(u_n=1|x_n=0)(p({S_{n-1}^1}'|{u_n=1,x_n=0}){}\nonumber\\&&
+p(u_n=0|x_n=1)(p({S_{n-1}^2}|{u_n=0,x_n=1}){}\nonumber\\&&+p(u_n=1|x_n=1)(p({S_{n-1}^2}'|{u_n=1,x_n=1}) \bigg) \; .
%\end{split}
\end{eqnarray}
And the winning probability of $S_{n-1}^j$, can be rewritten as,
\begin{eqnarray}
\label{temp2}
p({S_{n-1}^i}) &&= \frac{1}{2} \bigg( p(u_n=0|x_n=0)(p(S_{n-1}^j|{u_n=0,x_n=0}){}\nonumber\\&&+p(u_n=1|x_n=0)(p({S_{n-1}^j}|{u_n=1,x_n=0}){}\nonumber\\&&
+p(u_n=0|x_n=1)(p({S_{n-1}^j}|{u_n=0,x_n=1}){}\nonumber\\&&+p(u_n=1|x_n=1)(p({S_{n-1}^j}|{u_n=1,x_n=1}) \bigg) \; .
\end{eqnarray}
This way of expressing $p(S_{n-1}^j)$ is rather redundant. However, its presented here, for the sake of the observation that $p(S_{n-1}^j)$, does not depend on the input or output bits of the $n$th party. Now, it can be argued that the game $S_{n-1}
^j$ should be equivalent to one of the four $S_{n-1}$ games, $\{S_{n-1}^1,{S_{n-1}^1}',S_{n-1}^2,{S_{n-1}^2}'\}$ to get the overall maximum. W.l.o.g. we take $S_{n-1}^j \equiv S_n^1$. Now to determine $\max_{\mathcal{NS}}\{p({S_{n-1}^l})\}$ we consider only the case when $p({S_n^i}) \ge \frac{1}{2}$ ( as if $p({S_n^i}) \le \frac{1}{2}$, then we can consider $p({{S_n^i}{'}}) \ge \frac{1}{2}$.  We consider the following two sub-cases,
\begin{enumerate}
\item When $\frac{1}{2} \le p({S_n^i}) \le \frac{3}{4}$, $\max_{\mathcal{NS}}\{p({S_{n-1}^1})\}$. Consider the probability distribution for $n-1$ parties under the no-signaling condition that wins $S_{n-1}^1 \equiv S_{n}^i||(u_n=0,x_n=0)$ game with probability 1 that is $p({S_{n-1}^1})=1$. By using complementarity relation among $S_{n-1}$ games given by (\ref{marginal_game}) and using (\ref{temp2}), it can be seen that for such a probability distribution, $p({S_{n}^2})=p({S_{n}^2}')=\frac{1}{2}$ and by \textbf{Property \ref{trivial_complementary}.}, $p({S_{n}^1}')=0$. Now, if $p(u_n=0,x_n=0)=p(u_n=0,x_n=1)=\frac{1}{2}$ then by using (\ref{temp}) we obtain, $p({S_n^i})=\frac{1}{2}$ and if  $p(u_n=0,x_n=0)=p(u_n=0,x_n=1)=1$, then $p({S_n^i})=\frac{3}{4}$. Similarly for any value of $p({S_n^i})$ in the range $[\frac{1}{2},\frac{3}{4}]$, we can get $\max_{\mathcal{NS}}\{p({S_{n-1}^1})\}=1$ by taking the aforementioned mentioned 
probability distribution for the first $n-1$ parties and varying the marginals of the $n^{\text{th}}$ party.
\item When $\frac{3}{4} \le p({S_n^i}) \le 1$, upon comparing (\ref{temp}) and (\ref{temp2}) and using complementarity relations for the $S_{n-1}$ games given by (\ref{marginal_game}), we find $\max_{\mathcal{NS}}\{p(S_{n-1}^1)\}$ to be,
\begin{eqnarray}
\label{general_win_probability23}
%\begin{eqnaraay}
\max_{\mathcal{NS}}\{p(S_{n-1}^1)\} &&= \frac{1}{2} \bigg( p(u_n=0|x_n=0)p(S_{n-1}^1|{u_n=0,x_n=0}){}\nonumber\\&&+p(u_n=1|x_n=0)(1-p({S_{n-1}^1}'|{u_n=0,x_n=0})){}\nonumber\\&&
+p(u_n=0|x_n=1)(\frac{3}{2}-p(S_{n-1}^2|{u_n=0,x_n=0})){}\nonumber\\&&+p(u_n=1|x_n=1)(\frac{3}{2}-p({S_{n-1}^2}'|{u_n=0,x_n=0})) \bigg) \; ,
%\end{split}
\end{eqnarray}
which using the assignment the assignment $p({S_{n-1}^1}|{u_n=0,x_n=0})=1$ becomes,
\begin{eqnarray}
\label{general_win_probability3}
%\begin{eqnaraay}
\max_{\mathcal{NS}}\{p({S_{n-1}^1})\} &&= \frac{1}{2} \bigg( p(u_n=0|x_n=0)-2p({S_{n}^i})+\frac{5}{2}\bigg) \; .
%\end{split}
\end{eqnarray}
Using (\ref{chsh_marginal_result}), we get,
\begin{equation}
\max_{\mathcal{NS}}\{p({S_{n-1}^1})\}= \frac{1}{2} \bigg( \frac{5-4p({S_{n}^i})}{2}-2p({S_{n}^i})+\frac{5}{2}\bigg) \; ,
\end{equation}
\begin{equation}
\max_{\mathcal{NS}}\{p({S_{n-1}^j})\} =  \bigg( \frac{5-4p({S_{n}^i})}{2} \bigg) \; .
\end{equation}

\end{enumerate}
\subsection*{$S_{k}$ vs. $S_n$}

Now, we continue with the proof of the relation in \textbf{Theorem 2.} for any $k < n-1$. We find the value of $\max_{\mathcal{NS}}\{p(S_{k}^j)\}$, for a given value of $p({S_n^i})\ge \frac{1}{2}$. The winning probability of the $S_n^i$ game can be expressed as in (\ref{general_win_probability2}). And the winning probability of a $S_k^j$ game can be re-written as,
\begin{eqnarray}
\label{general_win_probability2}
&&p({S_k^j}) ={}\nonumber\\&& \frac{1}{2} \bigg( p(\mathcal{A}_{n-k}=0|\mathcal{B}_{n-k}=0)p({S_k^j}|\mathcal{A}_{n-k}=0,\mathcal{B}_{n-k}=0){}\nonumber\\&&+p(\mathcal{A}_{n-k}=1|\mathcal{B}_{n-k}=0)p({S_k^j}|\mathcal{A}_{n-k}=1|\mathcal{B}_{n-k}=0) \nonumber\\&&
+p(\mathcal{A}_{n-k}=0|\mathcal{B}_{n-k}=1)p({S_k^j}|\mathcal{A}_{n-k}=0,\mathcal{B}_{n-k}=1){}\nonumber\\&&+p(\mathcal{A}_{n-k}=1|\mathcal{B}_{n-k}=1)p({S_k^j}|\mathcal{A}_{n-k}=1,\mathcal{B}_{n-k}=1) \bigg) \;.{}\nonumber\\&& 
\end{eqnarray}
This way of expressing $p(S_{n-1}^j)$ is rather redundant. However, its presented here, for the sake of the observation that $p(S_{j}^j)$, does not depend on the input or output bits of the $(n-k)$ parties. Now, in-order to get the overall maximum $p(S_{j}^j)$, $S_k^j \in \{ S_k^1,{S_k^1}',S_k^2,{S_k^2}'\}$ where $S_k^1\equiv S_n^i||(\mathsf{A}_{n-k}=0,\mathsf{B}_{n-k}=0),{S_k^1}'\equiv S_n^i||(\mathsf{A}_{n-k}=1,\mathsf{B}_{n-k}=0),S_k^2\equiv S_n^i||(\mathsf{A}_{n-k}=0,\mathsf{B}_{n-k}=1),{S_k^2}'\equiv S_n^i||(\mathsf{A}_{n-k}=1,\mathsf{B}_{n-k}=1)$. We consider the following two sub cases:
\begin{enumerate}
\item When $\frac{1}{2} \le p_{win}^{S_n^i} \le \frac{3}{4}$, $\max_{\mathcal{NS}}\{(p(S_{k}^j)\}=1$. Consider the probability distribution for $k$ parties under no-signaling that wins the $S_{k}^1$ game with probability 1, i.e., $p(S_{k}^1)=1$. By using complementarity relation among $S_k$ games given in \textbf{Theorem 1}, it can be seen that for such a probability distribution $p(S_{k}^2)=p({S_{k}^2}')=\frac{1}{2}$ and by \textbf{Property \ref{trivial_complementary}.}, $p({S_{k}^1}')=0$. Now, taking $p(\mathcal{A}_{n-k}=0|\mathcal{B}_{n-k}=0)=p(\mathcal{A}_{n-k}=0|\mathcal{B}_{n-k}=1)=\frac{1}{2}$, we obtain  $p({S_n^i})=\frac{1}{2}$. Fixing,  $p(\mathcal{A}_{n-k}=0|\mathcal{B}_{n-k}=0)=p(\mathcal{A}_{n-k}=0|\mathcal{B}_{n-k}=1)=1$, then $p({S_n^i})=\frac{3}{4}$. Similarly, for all values of $p({S_n^i}) \in [\frac{1}{2},\frac{3}{4}]$ we can get $\max_{\mathcal{NS}}\{p(S_{k}^j)\}=1$. 
\item When $\frac{3}{4} \le p_{win}^{S_n^i} \le 1$. Using \textbf{Theorem 1.} we obtain,
\begin{eqnarray}
\label{sn_vs_sm_proof}
%\begin{eqnaraay}
\max_{\mathcal{NS}}\{p({S_{k}^j})\} &&= \frac{1}{2} \bigg( p(\mathcal{A}_{n-k}=0|\mathcal{B}_{n-k}=0)(p(S_{k}^1|\mathcal{A}_{n-k}=0,\mathcal{B}_{n-k}=0){}\nonumber\\&&+p(\mathcal{A}_{n-k}=1|\mathcal{B}_{n-k}=0)(1-p({S_{k}^1}'|\mathcal{A}_{n-k}=1,\mathcal{B}_{n-k}=0)\nonumber\\&&
+p(\mathcal{A}_{n-k}=0|\mathcal{B}_{n-k}=1)(\frac{3}{2}-p(S_{k}^2|\mathcal{A}_{n-k}=0,\mathcal{B}_{n-k}=1)){}\nonumber\\&&+p(\mathcal{A}_{n-k}=1|\mathcal{B}_{n-k}=1)(\frac{3}{2}-p({S_{k}^2}'|\mathcal{A}_{n-k}=1,\mathcal{B}_{n-k}=1) \bigg) \; . 
%\end{split}
\end{eqnarray}
Using the assignment $p(S_{k}^1|\mathcal{A}_{n-k}=0,\mathcal{B}_{n-k}=0)=1$, then (\ref{sn_vs_sm_proof}) reduces to,
%\begin{widetext}
\begin{eqnarray}
\label{general_win_probability3}
%\begin{eqnaraay}
\max_{\mathcal{NS}}\{p(S_{k}^1)\} &&= \frac{1}{2} \bigg( p(\mathcal{A}_{n-k}=0|\mathcal{B}_{n-k}=0)-2p({S_{n}})+\frac{5}{2}\bigg) \; . 
%\end{split}
\end{eqnarray}
Now we use complementarity relations between $S_n$ games and auxiliary games (see Appendix D) given by (\ref{chsh_marginal_result}) to get,
\begin{eqnarray}
&&\max_{\mathcal{NS}}\{p({S_{k}^1}\}= {}\nonumber\\&&
\frac{1}{2} \bigg( \frac{5-4p(S_{n}^i)}{2}-2p({S_{n}^i})+\frac{5}{2}\bigg) \; ,
\end{eqnarray}
\begin{eqnarray}
\label{sn_vs_sm_final}
\max_{\mathcal{NS}}\{p({S_k^j}\} \}=\bigg( \frac{5-4p({S_{n}^i})}{2} \bigg) \; .
\end{eqnarray}
\end{enumerate}

\end{appendices}

\end{widetext}